\documentclass[10pt,prd,twocolumn,showpacs,amsmath,amssymb,aps,floats,floatfix,nofootinbib]{revtex4-1}
\usepackage{amsmath,amsfonts,amssymb}
\usepackage{graphicx}
\usepackage{color}
\usepackage[dvipsnames,svgnames,x11names]{xcolor}
\usepackage{slashed} 
\usepackage{url}
\usepackage{subfigure}
\usepackage{multirow} 

\def\ifb{\mathrm{fb}^{-1}} 
\def\TeV{\mathrm{TeV}} 
\def\GeV{\mathrm{GeV}} 
\def\pT{p_\mathrm{T}} 
\def\missET{\slashed E_\mathrm{T}} 
\def\missE{\slashed E} 

\makeatother

\allowdisplaybreaks 

\begin{document}

\title{Dark matter searches in the mono-$Z$ channel at high energy $e^+e^-$ colliders}
\author{Zhao-Huan Yu$^1$}
\author{Xiao-Jun Bi$^1$}
\author{Qi-Shu Yan$^{2,3}$}
\author{Peng-Fei Yin$^1$}
\affiliation{$^1$Key Laboratory of Particle Astrophysics,
Institute of High Energy Physics, Chinese Academy of Sciences,
Beijing 100049, China}
\affiliation{$^2$School of Physics,
University of Chinese Academy of Sciences,
Beijing 100049, China}
\affiliation{$^3$Center for High Energy Physics, Peking University, Beijing 100871, China}

\begin{abstract}

We explore the mono-$Z$ signature for dark matter searches at future high energy $e^+e^-$ colliders. In the context of effective field theory, we consider two kinds of contact operators describing dark matter interactions with electroweak gauge bosons and with electron/positron, respectively. For five benchmark models, we propose kinematic cuts to distinguish signals from backgrounds for both charged leptonic and hadronic decay modes of the $Z$ boson. We also present the experimental sensitivity to cutoff scales of effective operators and compare it with that of the Fermi-LAT indirect search and demonstrate the gains in significance for the several configurations of polarized beams.

\end{abstract}


\maketitle


\section{Introduction}

Astrophysical and cosmological observations have established that dark matter (DM) is one of the main components of our universe. But the nature of DM is still a mystery. One kind of well motivated candidates for DM is the so-called weakly interacting massive particles (WIMPs), which can be produced from the thermal bath in the early universe. In order to explain the correct DM relic density observed today, WIMPs should interact with standard model (SM) particles via non-negligible couplings. Such interactions are expected to be testable at collider experiments, which are complementary to direct and indirect searches for DM.

DM particles can be produced directly or as decay products of other new particles at colliders. Once produced, DM particles will escape from the detector without energy deposit, like neutrinos. Therefore a large missing energy ($\missE$) is the typical signature of DM particles, which has been used to search for particles in supersymmetric models with R-parity. In order to determine the missing energy, other energetic objects associating with DM particles are required. Such energetic objects are useful handles and can be jets, leptons, photons, etc. When DM particles are directly produced or their mother particles are almost degenerate with them in mass, a better approach to probe the DM signal is to observe the excess in the $\text{mono-}X + \missE$ processes, where $X$ can be a jet~\cite{Beltran:2010ww,Goodman:2010yf,Bai:2010hh,
Ajaib:2011hs,Drees:2012dd,Yu:2012kj},
a photon~\cite{Fargion:1995qb,Fox:2011pm},
a $W$ boson~\cite{Bai:2012xg},
a $Z$ boson~\cite{Bell:2012rg,Carpenter:2012rg},
or a Higgs boson~\cite{Petrov:2013nia,Carpenter:2013xra}.

Typically, the interactions between DM and SM particles are model-dependent. In order to capture the most generic features of DM searches at colliders, it is helpful to use
the effective field theory approach~\cite{Beltran:2010ww,Goodman:2010yf,
Bai:2010hh,Fox:2011pm,Bai:2012xg,Carpenter:2012rg,
Carpenter:2013xra,Cao:2009uw,Rajaraman:2011wf,
Cheung:2012gi,Fox:2012ee,Cotta:2012nj,Ding:2013nvx,
Lin:2013sca,Nelson:2013pqa,Bell:2013wua,Lopez:2014qja,Fox:2011fx,
Dreiner:2012xm,Chae:2012bq,Yu:2013aca}.
In this work we will use contact operators to describe the interactions between DM and SM particles in a rather model-independent way.
Studies on DM signatures at hadron colliders have been performed by many authors~\cite{Beltran:2010ww,
Goodman:2010yf,Bai:2010hh,Fox:2011pm,Bai:2012xg,
Bell:2012rg,Carpenter:2012rg,Petrov:2013nia,
Carpenter:2013xra,Cao:2009uw,Rajaraman:2011wf,
Cheung:2012gi,Fox:2012ee,Cotta:2012nj,Ding:2013nvx,
Lin:2013sca,Nelson:2013pqa,Bell:2013wua,Lopez:2014qja,
Kamenik:2011nb,An:2012va,Tsai:2013bt,Chang:2013oia,An:2013xka,
Bai:2013iqa,DiFranzo:2013vra,deSimone:2014pda,Bai:2014osa,
Gomez:1404.1918,Izaguirre:1404.2018}.
On the other hand, future high energy $e^+e^-$ colliders, such as ILC, CEPC, and TLEP, can also be good places to probe
DM signatures~\cite{Fargion:1995qb,Fox:2011fx,Dreiner:2012xm,
Chae:2012bq,Yu:2013aca,Birkedal:2004xn,Konar:2009ae,Bartels:2012ex}.

The $\text{mono-}Z + \missE$ signature at the LHC has been considered in Ref.~\cite{Bell:2012rg,Carpenter:2012rg},
and an experimental analysis at $\sqrt{s}=8~\TeV$ with an integrated luminosity of $20.3~\ifb$ has been recently released
by the ATLAS Collaboration~\cite{Aad:2014vka}.
However, this signature at $e^+e^-$ colliders has not been done. In this work, we study the $\text{mono-}Z + \missE$ signature at future $e^+e^-$ colliders to fill this gap.

It is worthy to point out that the mono-$Z$ searching channel is sensitive to the DM interaction with $e^\pm$ as well as the DM interaction with $ZZ$/$Z\gamma$: 1) For the DM-$e^\pm$ interaction, which is difficult to be explored at hadron colliders, the mono-$Z$ boson is produced by the initial state radiation of $e^\pm$ beams. 2) For the DM interaction with $ZZ$/$Z\gamma$, a $Z$ boson and a pair of DM particles are directly produced via the $e^+ e^- \to Z^*/\gamma^* \to Z \chi \chi$ process. In this work, we consider two kinds of contact operators to describe these interactions. It is remarkable that compared with the situation at hadron colliders, the mono-$Z$ bosons can be reconstructed through both their charged leptonic and hadronic decay channels in a much cleaner environment. Another remarkable fact is that the full 4-momentum of the missing energy can be reconstructed.

This paper is organized as follows. In Sec.~\ref{sec:eff_op}, we introduce effective operators to describe the DM interactions with $e^\pm$ and with electroweak gauge bosons. In Sec.~\ref{sec:sim}, we present the simulation procedure for generating signal and background samples. We investigate both the charged leptonic and hadronic decay channels of $Z$ bosons. In Sec.~\ref{sec:sensi}, we estimate the experimental sensitivity to the interactions. We also compare the expected sensitivity at $e^+e^-$ colliders with the constraints from DM indirect searches. In Sec.~\ref{sec:pol}, we discuss the benefits of searching with polarized beams, which will be available at the ILC. The final section is our conclusions and discussions.

\section{Effective operators}
\label{sec:eff_op}

The mono-$Z$ channel at $e^+e^-$ colliders is sensitive to
the DM production process $e^+e^-\to\chi\bar\chi Z$.
As shown in Fig.~\ref{fig:d_DMDM_Z:a},
this process can be contributed by $s$-channel Feynman diagrams
due to the DM coupling to $ZZ/Z\gamma$.
Meanwhile, the DM coupling to $e^\pm$ can be another source of this process,
where the $Z$ boson comes from the initial state radiation,
as illustrated in Fig.~\ref{fig:d_DMDM_Z:b}.

\begin{figure*}[!htbp]
\centering
\subfigure[~Due to the DM coupling to $ZZ/Z\gamma$.\label{fig:d_DMDM_Z:a}]
{\includegraphics[height=.2\textwidth]{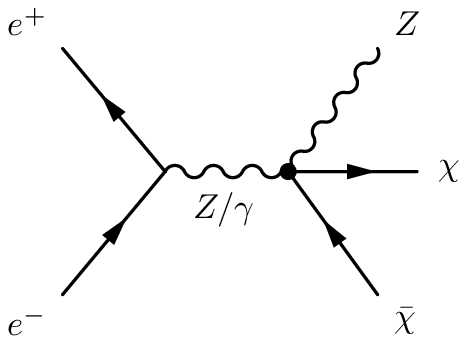}}
\hspace*{4em}
\subfigure[~Due to the DM coupling to $e^\pm$.\label{fig:d_DMDM_Z:b}]
{\includegraphics[height=.2\textwidth]{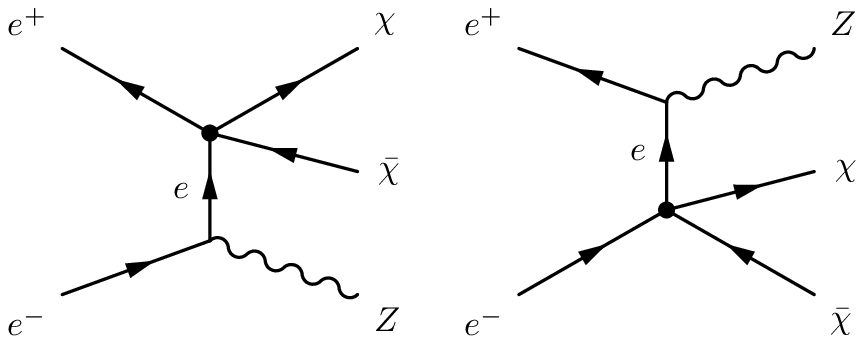}}
\caption{DM production processes $e^+e^-\to\chi\bar\chi Z$ are illustrated.}
\label{fig:d_DMDM_Z}
\end{figure*}

In this work, we assume that the DM particle is a Dirac fermion $\chi$
and a singlet under SM gauge interactions.
Effective operators are used to describe the interactions between DM and SM particles.
We adopt several representative operators to demonstrate the experimental sensitivity.

Comprehensive lists of effective operators for the interactions between DM and electroweak sectors
can be found in Refs.~\cite{Rajaraman:2012fu,Chen:2013gya}.
Regarding the $SU(2)_L\times U(1)_Y$ gauge invariance and $CP$ conservation,
a pair of DM particles may couple to a pair of gauge bosons
through either of the following dimension-7 operators:
\begin{eqnarray}
\mathcal{O}_\mathrm{F1} &=& \frac{1}{\Lambda_1^3}\bar \chi \chi B_{\mu\nu}B^{\mu\nu}
+ \frac{1}{\Lambda_2^3}\bar \chi \chi W_{\mu\nu}^a W^{a\mu\nu}, \label{op1}
\\
\mathcal{O}_\mathrm{F2} &=& \frac{1}{\Lambda_1^3}\bar \chi i{\gamma _5}\chi
B_{\mu\nu}{\tilde B}^{\mu\nu} + \frac{1}{\Lambda_2^3}\bar \chi i{\gamma _5}\chi
W_{\mu\nu}^a{\tilde W}^{a\mu\nu}, \label{op2}
\end{eqnarray}
where $B_{\mu\nu}$ and $W_{\mu\nu}^a$ are the field strength tensors
for the $U(1)_Y$ and $SU(2)_L$ gauge groups, respectively.
In terms of physical fields, we have
\begin{eqnarray}
\mathcal{O}_\mathrm{F1} &\supset& G_\mathrm{ZZ} \bar\chi\chi Z_{\mu\nu} Z^{\mu\nu}
+ G_\mathrm{AZ} \bar\chi\chi A_{\mu\nu} Z^{\mu\nu},
\\
\mathcal{O}_\mathrm{F2} &\supset& G_\mathrm{ZZ} \bar\chi i\gamma_5\chi Z_{\mu\nu} {\tilde Z}^{\mu\nu}
+ G_\mathrm{AZ} \bar\chi i\gamma_5\chi A_{\mu\nu} {\tilde Z}^{\mu\nu},
\end{eqnarray}
with
\begin{eqnarray}
G_\mathrm{ZZ} &\equiv& \dfrac{\sin^2\theta_W}{\Lambda_1^3}
+\dfrac{\cos^2\theta_W}{\Lambda_2^3},
\label{eq:G_ZZ}
\\
G_\mathrm{AZ} &\equiv& 2\sin\theta_W\cos\theta_W
\left(\dfrac{1}{\Lambda_2^3}-\dfrac{1}{\Lambda_1^3}\right).
\label{eq:G_AZ}
\end{eqnarray}
Here $\theta_W$ is the Weinberg angle,
and $A_{\mu\nu} \equiv \partial_\mu A_\nu - \partial_\nu A_\mu$
($Z_{\mu\nu} \equiv \partial_\mu Z_\nu - \partial_\nu Z_\mu$)
is the field strength tensor of the photon ($Z$ boson).
Through either $\mathcal{O}_\mathrm{F1}$ or $\mathcal{O}_\mathrm{F2}$,
DM particles generally couple to both $ZZ$ and $Z\gamma$.
However, if $\Lambda_1$ is equal to $\Lambda_2$, $G_\mathrm{AZ}$ would vanish
and the coupling to $Z\gamma$ would be turned off.

It is noticed that the DM coupling to $ZZ$ may also be originated from another
$SU(2)_L\times U(1)_Y$ gauge invariant dimension-7 operator~\cite{Carpenter:2012rg}
\begin{equation}
\mathcal{O}_\mathrm{FH} = \frac{1}{\Lambda^3}
\bar\chi\chi (D_\mu H)^\dag D_\mu  H.
\end{equation}
After the SM Higgs doublet $H$ acquires its vacuum expectation value,
this operator induces a dimension-5 operator
\begin{equation}
\frac{m_Z^2}{2\Lambda^3}\bar\chi\chi Z_\mu Z^\mu.
\end{equation}
At high energy colliders, the DM production resulted from this operator
is highly suppressed by a factor of $m_Z^4/s^2$
and the sensitivity could be poor. For the sake of comparison with the operators defined in Eq. (\ref{op1}) and Eq. (\ref{op2}),
we take into account this operator in the study.

The interaction between DM and $e^\pm$ can be described by
various effective operators~\cite{Zheng:2010js}.
We consider two dimension-6 operators as illuminating examples:
\begin{eqnarray}
\mathcal{O}_\mathrm{FP} &=& \frac{1}{\Lambda^2}
\bar \chi \gamma_5 \chi \bar e \gamma_5 e,
\\
\mathcal{O}_\mathrm{FA} &=& \frac{1}{\Lambda^2}
\bar\chi \gamma^\mu \gamma_5 \chi\bar e \gamma_\mu \gamma_5 e.
\end{eqnarray}
Although they have no essential difference in the dimensional analysis,
they exhibit distinct behaviors from each other
when polarized beams are considered, as we will see below.

\section{Numerical Analysis}
\label{sec:sim}

\subsection{Setup}

Here we briefly describe our simulation procedure
of generating background and signal samples
in the mono-$Z$ channel at $e^+e^-$ colliders.
According to $Z$ boson decay modes, we categorize the mono-$Z$ channel
into the charged leptonic channel and the hadronic channel, where the $Z$ boson
decays into two charged leptons and into two jets, respectively.

Both background and signal simulation samples are generated
by \texttt{MadGraph~5}~\cite{Alwall:2011uj}, where the model file is produced by
adding the DM particle and its effective couplings
to the SM through \texttt{FeynRules}~\cite{Alloul:2013bka}.
\texttt{PYTHIA~6}~\cite{Sjostrand:2006za} is used to carry out
parton shower, hadronization, and decay processes.
Fast detector simulation is performed by \texttt{PGS}~\cite{pgs}.
Jets are reconstructed by the anti-$k_\mathrm{T}$ clustering
algorithm~\cite{Cacciari:2008gp} with the jet parameter $R=0.4$.
From the technical design report of
ILC detectors~\cite{Behnke:2013lya},
the energy smearing parameters of the electromagnetic calorimeter
and of the hadronic calorimeter are implemented to the PGS card, which are
listed below as
\begin{equation}
\frac{\Delta E}{E}=\frac{17\%}{\sqrt{E/\GeV}}\oplus 1\%
\text{ and }
\frac{\Delta E}{E}=\frac{30\%}{\sqrt{E/\GeV}},
\end{equation}
respectively.

We consider three collision energies,
$\sqrt{s}=250~\GeV$, 500~GeV and 1~TeV,
which will be available at the ILC.
Future circular Higgs factories, like CEPC and TLEP,
may only have a collision energy of $\sim 240~\GeV$ or so.
Nevertheless, For the DM searching we study here,
their sensitivities would be similar to the case of $\sqrt{s}=250~\GeV$,
given the same integrated luminosity.

\begin{table}[!htbp]
\centering
\setlength\tabcolsep{0.3em}
\caption{For the case of $\sqrt{s}=500~\GeV$,
the benchmark points of effective operators are listed.
For the operators $\mathcal{O}_\mathrm{F1}$ and $\mathcal{O}_\mathrm{F2}$, $\Lambda=\Lambda_1=\Lambda_2$ is assumed.
$\sigma$ is the production cross section
of $e^+e^-\to\chi\bar\chi Z$ computed by \texttt{MadGraph}.}
\label{tab:bmp}
\begin{tabular}{cccc}
\hline\hline
 & $\Lambda$~(GeV) & $m_\chi$~(GeV) & $\sigma$~(fb) \\
\hline
$\mathcal{O}_\mathrm{F1}$ & 280 & 50 & 48.4 \\
$\mathcal{O}_\mathrm{F2}$ & 250 & 80 & 53.4 \\
$\mathcal{O}_\mathrm{FH}$ & 100 & 5 & 45.0 \\
$\mathcal{O}_\mathrm{FP}$ & 400 & 120 & 58.6 \\
$\mathcal{O}_\mathrm{FA}$ & 280 & 150 & 50.2 \\
\hline\hline
\end{tabular}
\end{table}
Below, we use the case of $\sqrt{s}=500~\GeV$ as an example
to demonstrate how we choose kinematic cuts.
For signals, the benchmark points we adopt are listed in Tab.~\ref{tab:bmp}.
Note that we deliberately choose low cutoff scales in the benchmark points so as to make
signals easily visable in distribution plots given in Fig. (\ref{fig:2lep_dist}) and Fig. (\ref{fig:2jet_dist}).

\subsection{Charged Leptonic channel}

In the charged leptonic channel ($\ell\ell+\missE$), the dominant SM background is
$e^+e^-\to\ell^+\ell^-\bar\nu\nu$ ($\ell=e,\mu$), which involves 2-body productions
$e^+ e^-\to W^+ W^-$ and $e^+ e^-\to ZZ$ with leptonic decays.
Minor backgrounds are $e^+e^-\to\tau^+\tau^-$ and $e^+e^-\to\tau^+\tau^-\bar\nu\nu$.
The selection cuts for three collision energies are summarized in Tab.~\ref{tab:2lep_cuts}.

\begin{table}[!htbp]
\centering
\setlength\tabcolsep{0.3em}
\caption{Selection cuts in the charged leptonic channel for three collision energies are shown.}
\label{tab:2lep_cuts}
\begin{tabular}{cccc}
\hline\hline
$\sqrt{s}$ & $250~\GeV$ & $500~\GeV$ & $1~\TeV$ \\
\hline
\multirow{2}{*}{Cut 1} & \multicolumn{3}{c}{2 OSSF leptons with $\pT>10~\GeV$, $|\eta|<3$} \\
& \multicolumn{3}{c}{No other particle or jet with $\pT>10~\GeV$, $|\eta|<3$} \\
Cut 2 & $\missET>15~\GeV$ & $\missET>30~\GeV$ & $\missET>40~\GeV$ \\
Cut 3 & \multicolumn{3}{c}{$|m_{\ell\ell}-m_Z|<5~\GeV$} \\
Cut 4 & $m_\mathrm{rec}\geq 100~\GeV$ & $m_\mathrm{rec}\geq 140~\GeV$ & $m_\mathrm{rec}\geq 200~\GeV$ \\
Cut 5 & / & \multicolumn{2}{c}{$25^\circ<\theta_{\ell\ell}<155^\circ$} \\
\hline\hline
\end{tabular}
\end{table}

Below we present a detailed explanation to these cuts in the case $\sqrt{s}=500~\GeV$ as an illustrating example.

\textit{Cut~1.}---Select the events containing
two opposite-sign same-flavor (OSSF)
electrons or muons with $\pT>10~\GeV$ and $|\eta|<3$.
Veto the events if there were any other hard lepton, tau, photon, or jet
with $\pT>10~\GeV$ and $|\eta|<3$.

This is the acceptance cut, which picks up the potential events with $Z$ boson leptonic decays.
It is justifiable to neglect taus which decay hadronically while taus which decay leptonically may be selected.
It is observed that quite a large amounts of the backgrounds $\tau^+\tau^-$
and $\tau^+\tau^-\bar\nu\nu$ can remain after cut~1.
Obviously, if the tau tagging for leptonical modes can work very well at $e^+e^-$ colliders, we can
use tau veto to further suppress such backgrounds.
The veto on other extra particles and jets in an event can reject other types of backgrounds,
such as $e^+e^-\to\ell\ell\gamma$, $e^+e^-\to\ell\ell\tau\tau$, or
$e^+e^-\to\tau\tau\tau\tau$, etc. Therefore we can safely neglect them in this study.
It should be pointed out that if the tau tagging for hadronic mode can work very well at $e^+ e^-$ colliders,
the hadronic taus can be also useful for reconstructing $Z$ bosons.

\begin{figure*}[!htbp]
\centering
\subfigure[~Normalized distributions of $\missET$ after cut~1.\label{fig:2lep_dist:mEt}]
{\includegraphics[width=.4\textwidth]{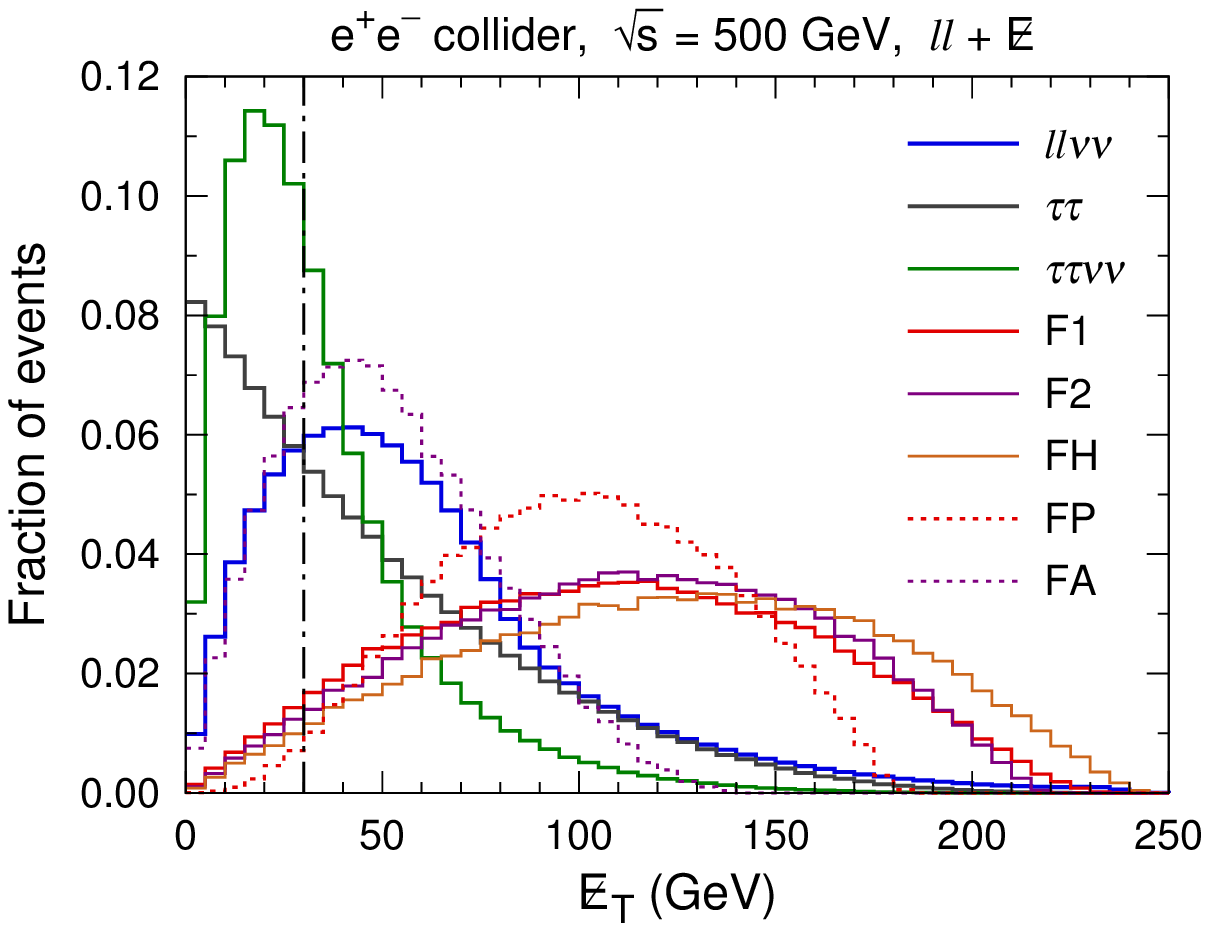}}
\subfigure[~Differential cross sections with respect to $m_{\ell\ell}$ after cut~2.\label{fig:2lep_dist:mll}]
{\includegraphics[width=.4\textwidth]{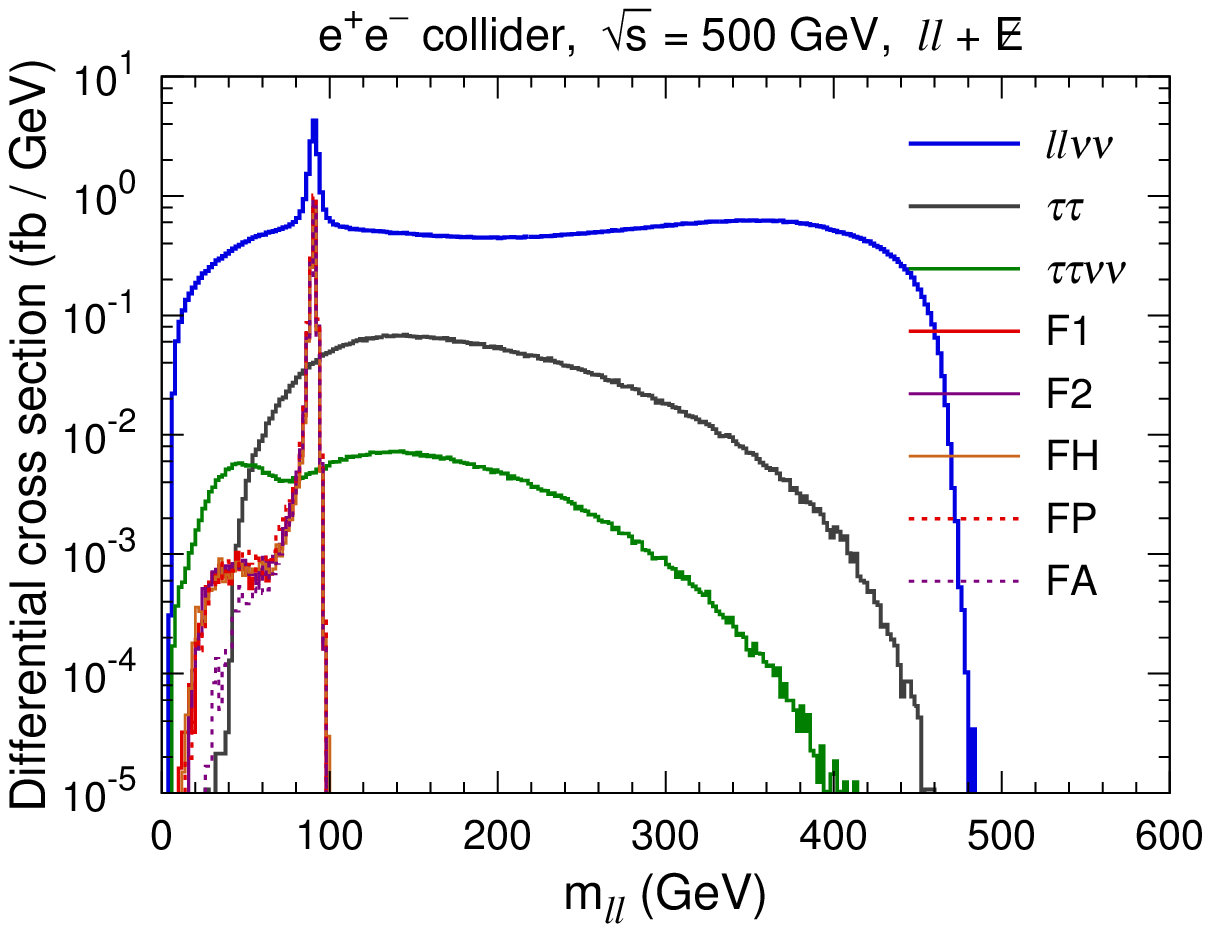}}
\subfigure[~Differential cross sections with respect to $m_\mathrm{rec}$ after cut~3.\label{fig:2lep_dist:mrec}]
{\includegraphics[width=.4\textwidth]{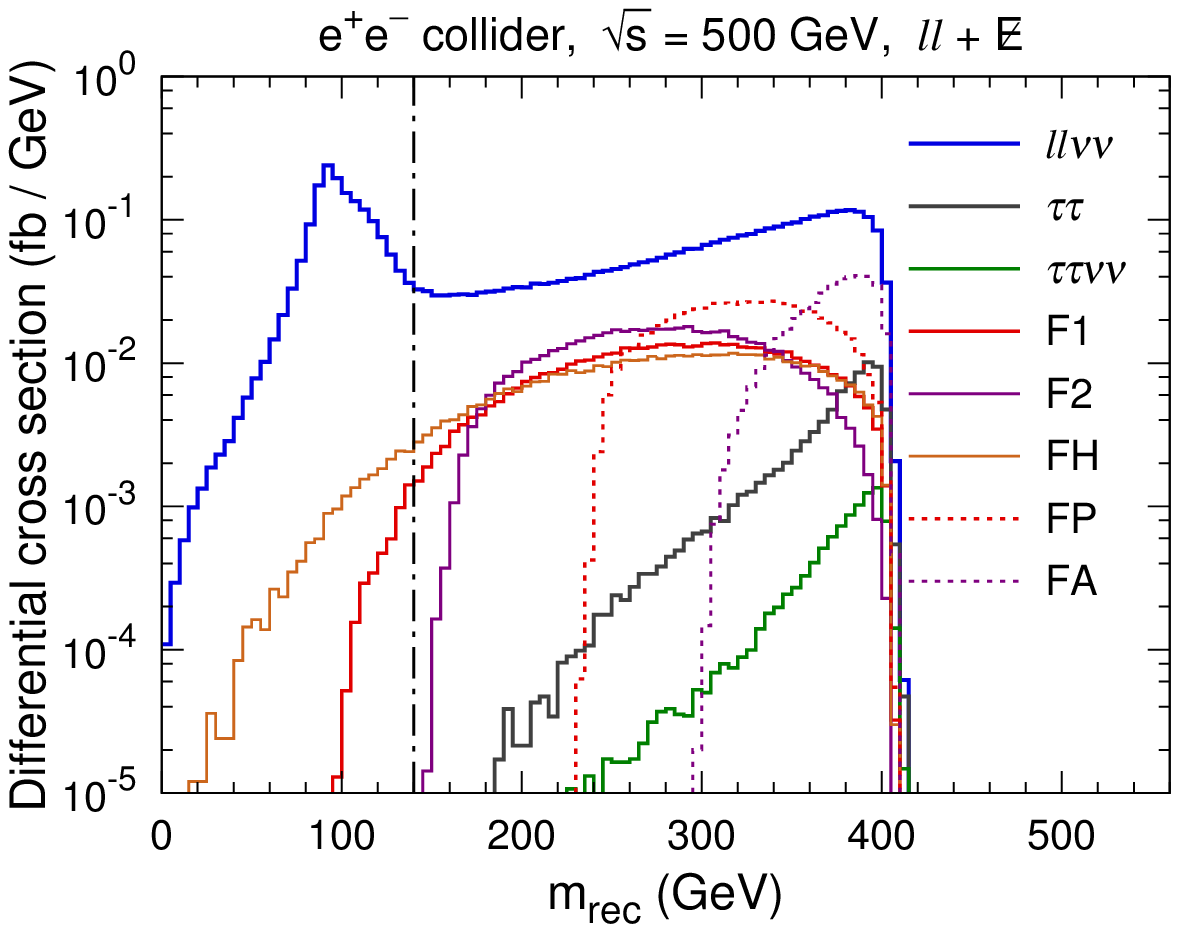}}
\subfigure[~Normalized distributions of $\theta_{\ell\ell}$ after cut~4.\label{fig:2lep_dist:thll}]
{\includegraphics[width=.4\textwidth]{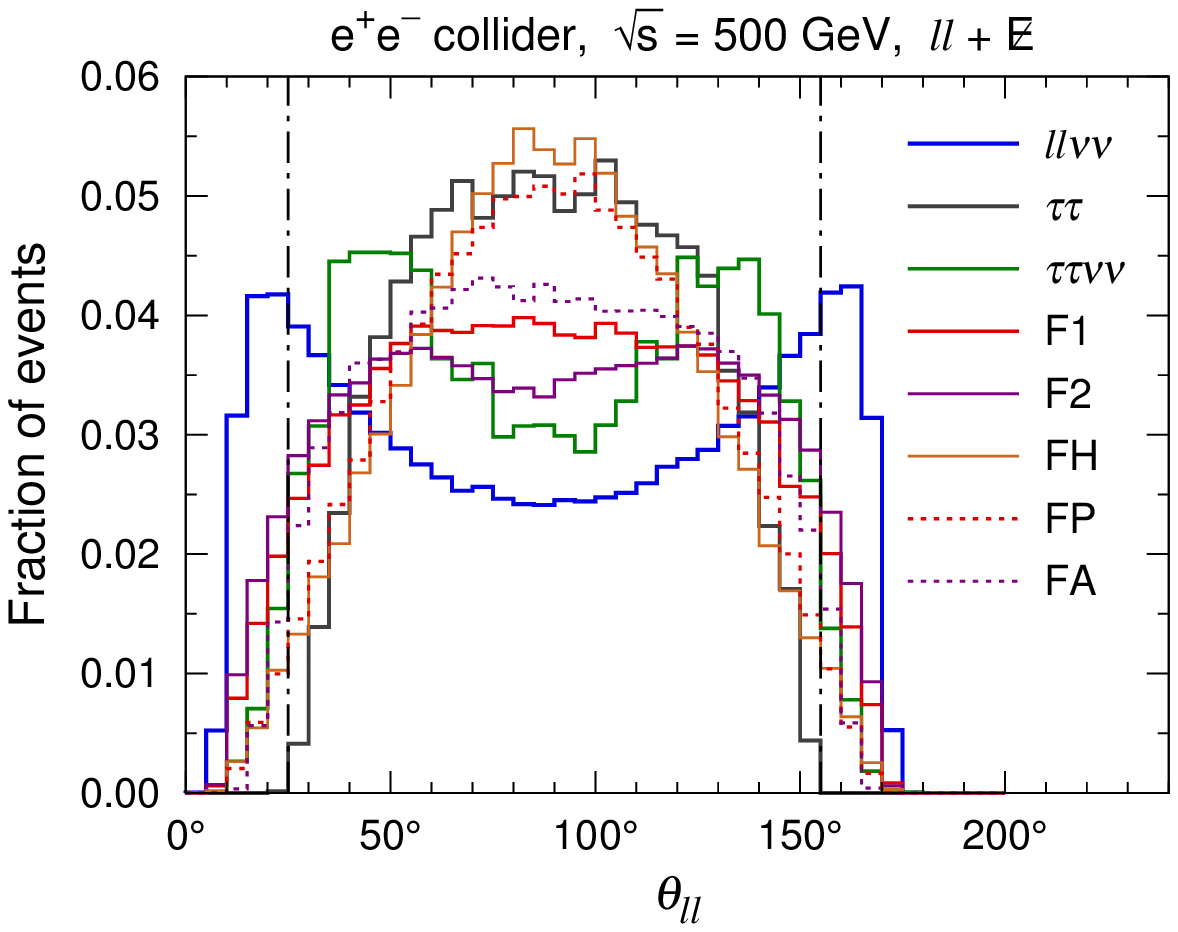}}
\caption{Distributions of backgrounds and signals in the charged leptonic channel for $\sqrt{s}=500~\GeV$ are demonstrated. The dot-dashed vertical lines denote the locations of our selection cuts. Note that the distributions in Figs.~\ref{fig:2lep_dist:mEt} and \ref{fig:2lep_dist:thll} are normalized, while those in Figs.~\ref{fig:2lep_dist:mll} and \ref{fig:2lep_dist:mrec} are not.}
\label{fig:2lep_dist}
\end{figure*}

\textit{Cut~2.}---Select the events with $\missET>30~\GeV$.

The normalized distributions of $\missET$ after cut~1
are displayed in Fig.~\ref{fig:2lep_dist:mEt}.
The backgrounds $\tau^+\tau^-$ and $\tau^+\tau^-\bar\nu\nu$
have basically more than four neutrinos in the final state,
and the missing momenta contributed by these neutrinos
would cancel out among themselves.
Therefore, these backgrounds tend to have small $\missET$.
This explains why cut~2 can suppress them.

\textit{Cut~3.}---Select the events where
the invariant mass of the two leptons
$m_{\ell\ell}$ satisfying $|m_{\ell\ell}-m_Z|<5~\GeV$.

Fig.~\ref{fig:2lep_dist:mll} shows the distributions
of $m_{\ell\ell}$ after cut~2.
The $Z$ boson peak is clearly demonstrated
in the distributions of the signals and
of the irreducible background $\ell^+\ell^-\bar\nu\nu$.
For the signals, there are some small bumps near $\sim 40~\GeV$,
which are due to leptonically decaying taus from $Z\to\tau^+\tau^-$.
Cut~3 picks up the events with on-shell leptonically decaying $Z$ bosons
and eliminates most of the events from the backgrounds
$\tau^+\tau^-$ and $\tau^+\tau^-\bar\nu\nu$.
The background $\ell^+\ell^-\bar\nu\nu$ from $e^+ e^- \to W^+ W^-$ is also highly suppressed.

\textit{Cut~4.}---Veto the events with $m_\mathrm{rec}<140~\GeV$.

The recoil mass against the reconstructed $Z$ boson
is defined as $m_\mathrm{rec}\equiv\sqrt{(p_{e^+}+p_{e^-}
-p_{\ell_1}-p_{\ell_2})^2}$, where $p_{e^+}$ ($p_{e^-}$)
is the 4-momentum of the initial positron (electron),
while $p_{\ell_1}$ and $p_{\ell_2}$ are the 4-momenta of the two leptons.
The background $\ell^+\ell^-\bar\nu\nu$ involves a process $e^+ e^- \to Z Z $ with two on-shell
$Z$ bosons, one of which decays into two leptons and the other one
decays into two neutrinos. As shown in Fig.~\ref{fig:2lep_dist:mrec},
this process leads to a peak around the $Z$ boson mass
in the $m_\mathrm{rec}$ distribution, and it can be removed by Cut~4.
It is remarkable that the recoil mass distributions of the signals
begin at $m_\mathrm{rec} = 2 m_\chi$, which may be useful for
determining the mass of the DM particle.

\textit{Cut~5.}---Select the events with
$25^\circ<\theta_{\ell\ell}<155^\circ$.

Here $\theta_{\ell\ell}$ is defined as the zenith angle of
the \textit{momentum sum} of the two leptons.
For the background $\ell^+\ell^-\bar\nu\nu$, the reconstructed $Z$ boson
may come from the initial state radiation and be close to the beam pipe,
as shown in Fig.~\ref{fig:2lep_dist:thll}. While for signal events, $Z$ bosons
mostly fly in the central regions.
Cut~5 can suppress this background to some degree.
Nonetheless, in order to keep more signal events, we are restricted from imposing a stricter cut.

\begin{table*}[!htbp]
\centering
\setlength\tabcolsep{0.4em}
\caption{Cross sections $\sigma$ (in fb) and signal significances $\mathcal{S}$
after each cut in the charged leptonic channel at $\sqrt{s}=500~\GeV$ are shown.
The significances are computed by assuming
an integrated luminosity $100~\ifb$.}
\label{tab:2lep_signif}
\begin{tabular}{cccccccccccccc}
\hline\hline
 & $\ell^+\ell^-\bar\nu\nu$ & $\tau^+\tau^-$ & $\tau^+\tau^-\bar\nu\nu$ & \multicolumn{2}{c}{$\mathcal{O}_\mathrm{F1}$} & \multicolumn{2}{c}{$\mathcal{O}_\mathrm{F2}$} & \multicolumn{2}{c}{$\mathcal{O}_\mathrm{FH}$} & \multicolumn{2}{c}{$\mathcal{O}_\mathrm{FP}$} & \multicolumn{2}{c}{$\mathcal{O}_\mathrm{FA}$} \\
 & $\sigma$ & $\sigma$ & $\sigma$ & $\sigma$ & $\mathcal{S}$ & $\sigma$ & $\mathcal{S}$ & $\sigma$ & $\mathcal{S}$ & $\sigma$ & $\mathcal{S}$ & $\sigma$ & $\mathcal{S}$ \\
\hline
Cut 1 & 306 & 20.4 & 2.85 & 2.65 & 1.46 & 2.94 & 1.61 & 2.47 & 1.36 & 3.24 & 1.78 & 2.86 & 1.57 \\
Cut 2 & 235 & 11.8 & 1.29 & 2.52 & 1.60 & 2.82 & 1.78 & 2.39 & 1.51 & 3.19 & 2.01 & 2.19 & 1.38 \\
Cut 3 & 23.9 & 0.410 & 0.0495 & 2.41 & 4.67 & 2.70 & 5.18 & 2.29 & 4.44 & 3.06 & 5.84 & 2.09 & 4.07 \\
Cut 4 & 16.0 & 0.410 & 0.0495 & 2.39 & 5.51 & 2.70 & 6.16 & 2.19 & 5.08 & 3.06 & 6.92 & 2.09 & 4.86 \\
Cut 5 & 12.1 & 0.410 & 0.0471 & 2.19 & 5.69 & 2.42 & 6.24 & 2.11 & 5.50 & 2.95 & 7.47 & 2.01 & 5.25 \\
\hline\hline
\end{tabular}
\end{table*}

In Tab.~\ref{tab:2lep_signif}, we tabulate the cross sections
of backgrounds and signals after each cut.
After applying cut~2 to 5, only $2\%-4\%$ of the background samples remain.
We define the signal significance as
$\mathcal{S}=S/\sqrt{S+B}$, where $B$ is the total number of
background events and $S$ is the number of signal events.
With an integrated luminosity of $100~\ifb$,
$\mathcal{S}$ for the benchmark points are also listed in Tab.~\ref{tab:2lep_signif},
from which we can see how the cuts improve the signal significance.

\subsection{Hadronic channel}

In the hadronic channel ($jj+\missE$), the dominant SM backgrounds are
$e^+ e^- \to jj\nu\bar\nu$ and $e^+ e^- \to jj\ell\nu$, which contain 2-body productions
$e^+ e^- \to Z Z$ and $e^+ e^- \to W^+ W^-$ with semileptonic decays, respectively.
Additionally, $e^+e^-\to t\bar t$ is a minor background.
The selection cuts for three collision energies are summarized in Tab.~\ref{tab:2jet_cuts}.

\begin{table*}[!htbp]
\centering
\setlength\tabcolsep{0.4em}
\caption{Selection cuts in the hadronic channel are shown.}
\label{tab:2jet_cuts}
\begin{tabular}{cccc}
\hline\hline
$\sqrt{s}$ & $250~\GeV$ & $500~\GeV$ & $1~\TeV$ \\
\hline
\multirow{2}{*}{Cut 1} & \multicolumn{3}{c}{Exact 2 jets with $\pT>10~\GeV$, $|\eta|<3$} \\
& \multicolumn{3}{c}{No other particle with $\pT>10~\GeV$, $|\eta|<3$} \\
Cut 2 & $\missET>15~\GeV$ & $\missET>30~\GeV$ & $\missET>40~\GeV$ \\
Cut 3 & $50~\GeV<m_{jj}<95~\GeV$ & \multicolumn{2}{c}{$40~\GeV<m_{jj}<95~\GeV$} \\
Cut 4 & $m_\mathrm{rec}\geq 120~\GeV$ & $m_\mathrm{rec}\geq 200~\GeV$ & $m_\mathrm{rec}\geq 300~\GeV$ \\
Cut 5 & $30^\circ<\theta_{jj}<150^\circ$ & \multicolumn{2}{c}{$25^\circ<\theta_{jj}<155^\circ$} \\
\hline\hline
\end{tabular}
\end{table*}

We expose the reasons for these cuts by using the case of $\sqrt{s}=500~\GeV$ as an example.

\textit{Cut~1.}---Select the events containing
two jets with $\pT>10~\GeV$ and $|\eta|<3$.
Veto the events if there were any other lepton, tau, photon, or jet
with $\pT>10~\GeV$ and $|\eta|<3$.

The two jets picked up by this acceptance cut will be used to reconstruct the hadronically decaying $Z$ boson.
It is realized that the $jj\nu\bar\nu$ is an irreducible background.
We also notice that there is a considerable amount of the background $jj\ell\nu$ remained after cut~1,
since the lepton $\ell$ can be clustered into a jets, or be with a low $\pT$,
or be close to undetectable regions, like the beam direction. On the other hand,
the background $t\bar t$ will be much less important, because it tends to
have either a large jet multiplicity or more than one energetic lepton.

\textit{Cut~2.}---Select the events with $\missET>30~\GeV$.

This cut is used to suppress the backgrounds $jj\nu\bar\nu$ and $jj\ell\nu$,
as shown in Fig.~\ref{fig:2jet_dist:mEt}.

\begin{figure*}[!htbp]
\centering
\subfigure[~Normalized distributions of $\missET$ after cut~1.\label{fig:2jet_dist:mEt}]
{\includegraphics[width=.4\textwidth]{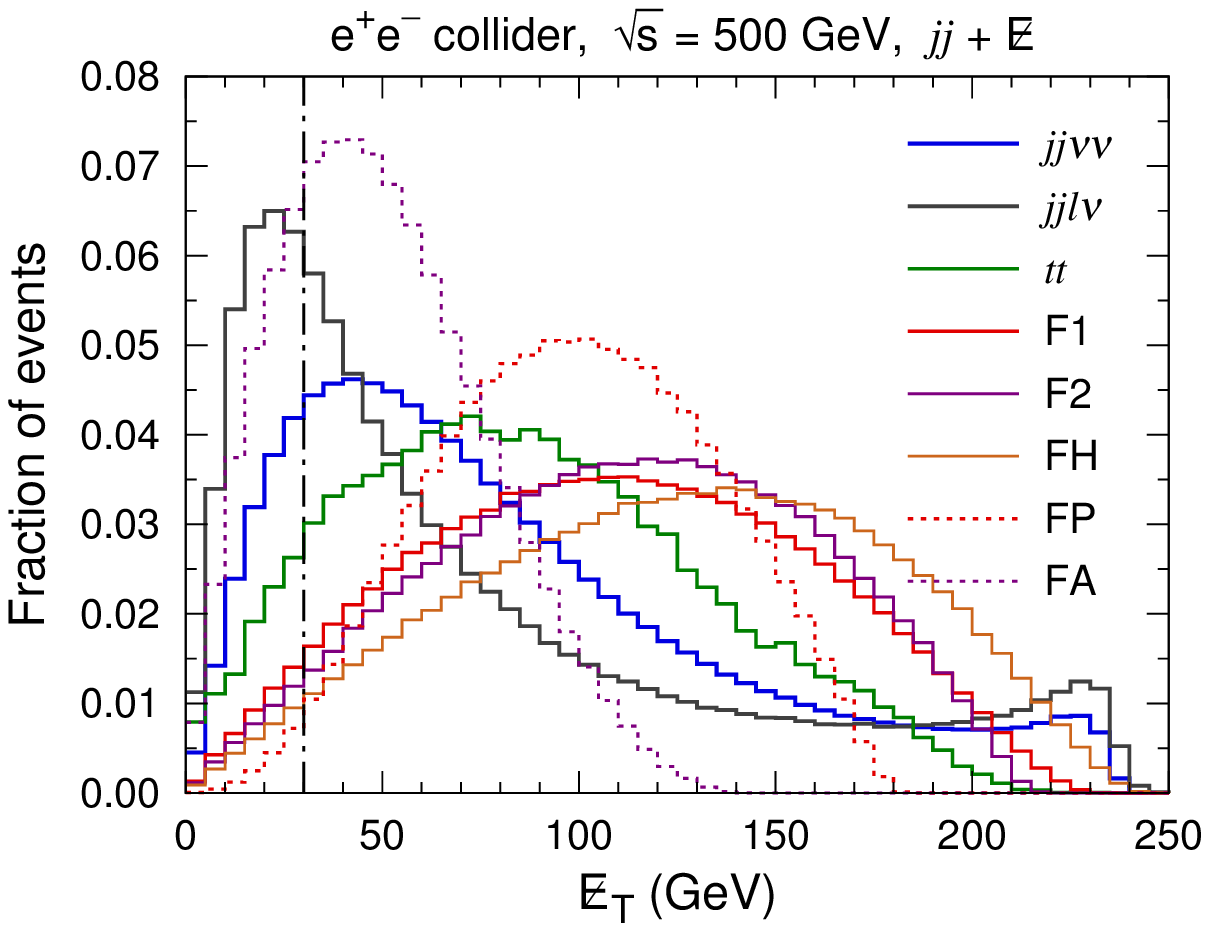}}
\subfigure[~Differential cross sections with respect to $m_{jj}$ after cut~2.\label{fig:2jet_dist:mjj}]
{\includegraphics[width=.4\textwidth]{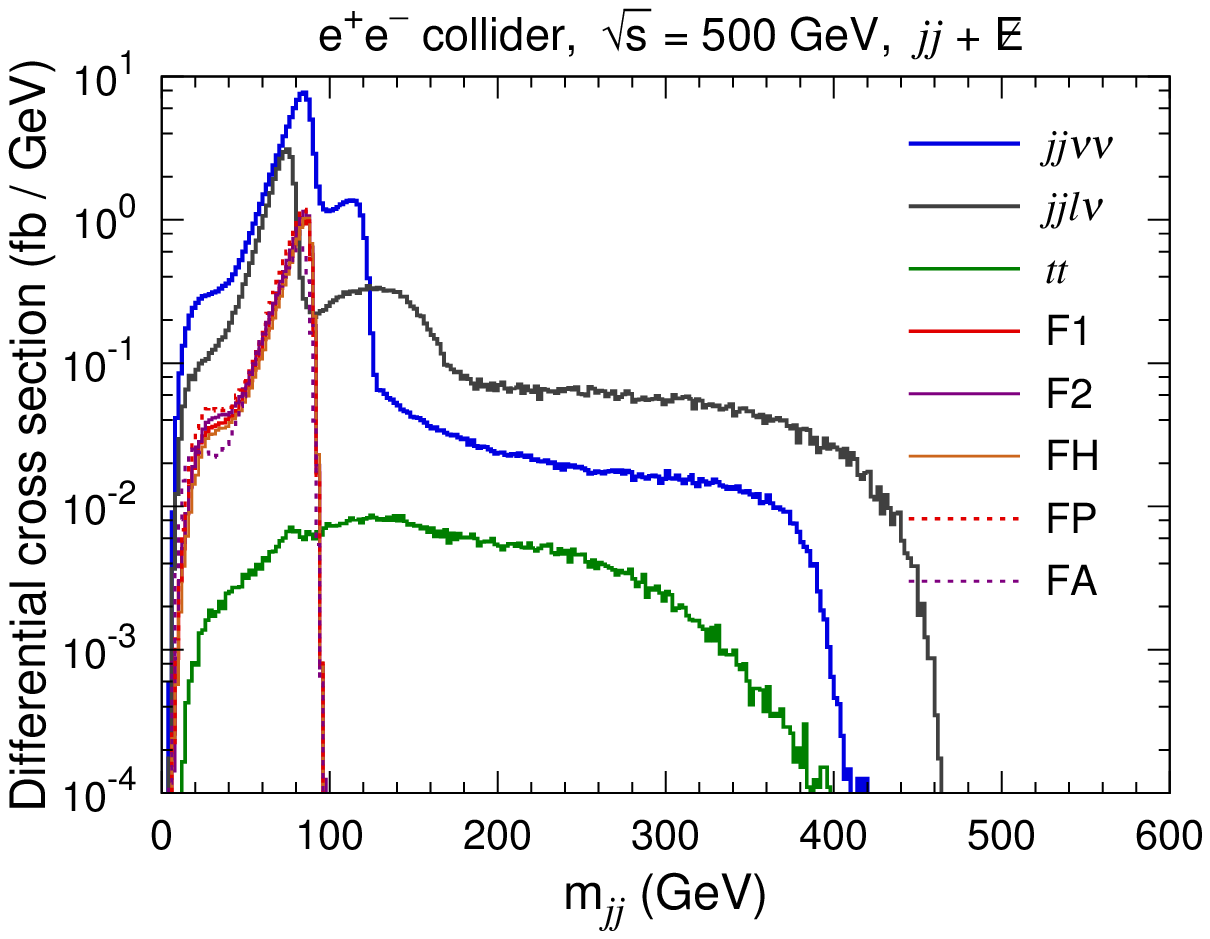}}
\subfigure[~Normalized distributions of $m_\mathrm{rec}$ after cut~3.\label{fig:2jet_dist:mrec}]
{\includegraphics[width=.4\textwidth]{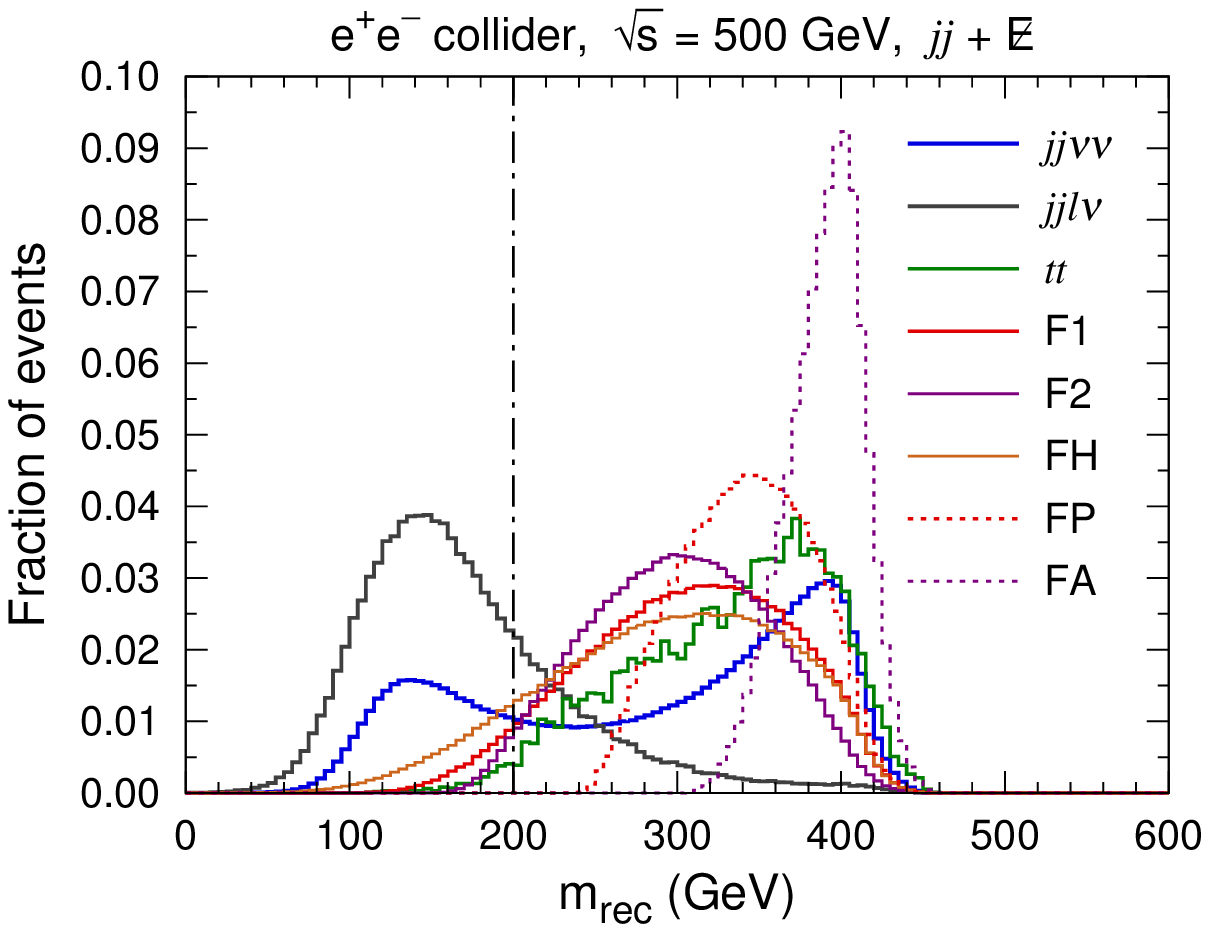}}
\subfigure[~Normalized distributions of $\theta_{jj}$ after cut~4.\label{fig:2jet_dist:thjj}]
{\includegraphics[width=.4\textwidth]{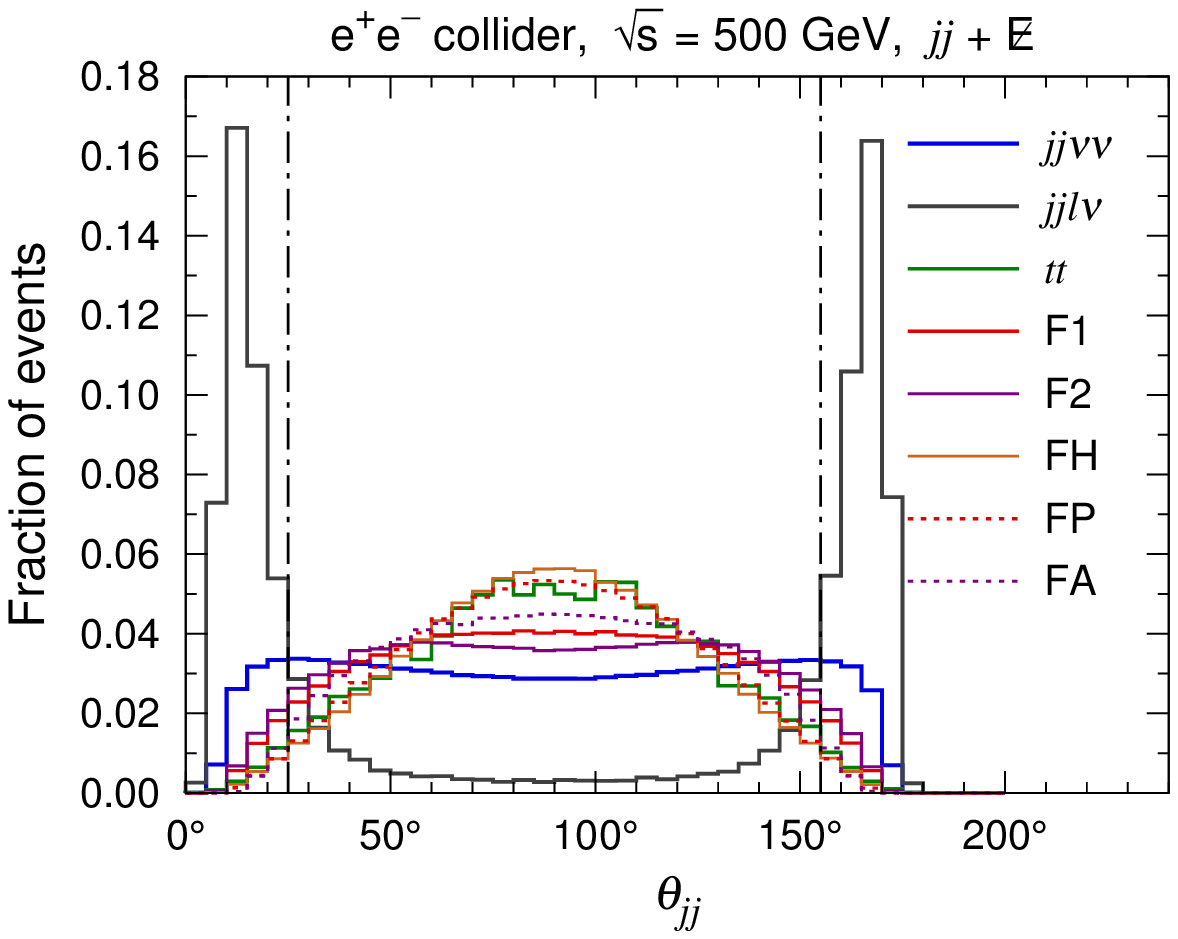}}
\caption{Distributions of backgrounds and signals in the hadronic channel for $\sqrt{s}=500~\GeV$ are demonstrated. The dot-dashed vertical lines denote the location of our selection cuts. Note that the distributions in Figs.~\ref{fig:2jet_dist:mEt}, \ref{fig:2jet_dist:mrec}, and \ref{fig:2lep_dist:thll} are normalized, while those in Fig.~\ref{fig:2jet_dist:mjj} are not.}
\label{fig:2jet_dist}
\end{figure*}

\textit{Cut~3.}---Select the events where
the invariant mass of the two jets
$m_{jj}$ satisfying $40~\GeV<m_{jj}<95~\GeV$.

After cut~2, the distributions of $m_{jj}$ are illustrated
in Fig.~\ref{fig:2jet_dist:mjj}.
The $Z$ peak is reconstructed at $\sim 85~\GeV$
for the signals and the background $jj\nu\bar\nu$,
while the $W$ peak are reconstructed at $\sim 75~\GeV$
for the background $jj\ell\nu$.
In addition, a second peak at $\sim 115~\GeV$ in the distribution
of $jj\nu\bar\nu$ comes from hadronically decaying Higgs bosons.
Due to the neutrinos from cascade decays of hadons,
the positions of reconstructed peaks are lower than the masses of the corresponding bosons.
Cut~3 is deliberately chosen to be loose to preserve most of the signals.
As a tradeoff, it also keeps a large number of events in the $W$ peak
from the background $jj\ell\nu$.

\textit{Cut~4.}---Veto the events with $m_\mathrm{rec}<200~\GeV$.

The recoil mass against the reconstructed $Z$ boson
is defined as $m_\mathrm{rec}\equiv\sqrt{(p_{e^+}+p_{e^-}
-p_{j_1}-p_{j_2})^2}$, where $p_{j_1}$ and $p_{j_2}$
are the 4-momenta of the two jets.
As demonstrated in Fig.~\ref{fig:2jet_dist:mrec},
the $m_\mathrm{rec}$ distributions of both the backgrounds $jj\nu\bar\nu$
and $jj\ell\nu$ has broad peaks at $\sim 140~\GeV$.
Cut~4 is useful to suppress these two types of backgrounds.

\textit{Cut~5.}---Select the events with
$25^\circ<\theta_{jj}<155^\circ$.

Here $\theta_{jj}$ is defined as the zenith angle of
the \textit{momentum sum} of the two jets.
As presented in Fig.~\ref{fig:2jet_dist:thjj},
this cut is very useful to suppress the background $jj\ell\nu$.
Obviously, if detectors can have better coverage in $\eta$, we can tag those forward hadronic $W$ bosons, which can be useful for rejecting the $j j \ell \nu$ background.
In comparison, the reconstructed $Z$ bosons in the background $jj\nu\bar\nu$ fly in a uniform way along the direction of $\eta$.

\begin{table*}[!htbp]
\centering
\setlength\tabcolsep{0.4em}
\caption{Cross sections $\sigma$ (in fb) and signal significances $\mathcal{S}$
after each cut in the hadronic channel at $\sqrt{s}=500~\GeV$ are shown.
The significances are computed by assuming
an integrated luminosity $100~\ifb$.}
\label{tab:2jet_signif}
\begin{tabular}{cccccccccccccc}
\hline\hline
 & $jj\nu\bar\nu$ & $jj\ell\nu$ & $t\bar t$ & \multicolumn{2}{c}{$\mathcal{O}_\mathrm{F1}$} & \multicolumn{2}{c}{$\mathcal{O}_\mathrm{F2}$} & \multicolumn{2}{c}{$\mathcal{O}_\mathrm{FH}$} & \multicolumn{2}{c}{$\mathcal{O}_\mathrm{FP}$} & \multicolumn{2}{c}{$\mathcal{O}_\mathrm{FA}$} \\
 & $\sigma$ & $\sigma$ & $\sigma$ & $\sigma$ & $\mathcal{S}$ & $\sigma$ & $\mathcal{S}$ & $\sigma$ & $\mathcal{S}$ & $\sigma$ & $\mathcal{S}$ & $\sigma$ & $\mathcal{S}$ \\
\hline
Cut 1 & 245 & 131 & 1.74 & 18.9 & 9.47 & 20.9 & 10.4 & 17.8 & 8.94 & 22.1 & 11.1 & 18.4 & 9.24 \\
Cut 2 & 207 & 93.2 & 1.56 & 18.0 & 10.0 & 20.0 & 11.2 & 17.2 & 9.64 & 21.8 & 12.1 & 13.9 & 7.84 \\
Cut 3 & 160 & 56.6 & 0.270 & 17.2 & 11.2 & 19.2 & 12.5 & 16.6 & 10.8 & 20.7 & 13.5 & 13.3 & 8.76 \\
Cut 4 & 115 & 14.9 & 0.264 & 16.3 & 13.4 & 18.7 & 15.3 & 14.6 & 12.1 & 20.7 & 16.9 & 13.3 & 11.1 \\
Cut 5 & 92.6 & 2.91 & 0.253 & 15.1 & 14.3 & 17.1 & 16.1 & 14.1 & 13.5 & 20.1 & 18.7 & 12.9 & 12.3 \\
\hline\hline
\end{tabular}
\end{table*}

We tabulate the cross sections of backgrounds and signals
and the signal significances after each cut in Tab.~\ref{tab:2jet_signif}.
After applying cut~2 to 5, $62\%$ of the irreducible background
$jj\nu\bar\nu$ are killed, while only $2\%$ of the background $jj\ell\nu$ remains.
Since the hadronic decay modes of the $Z$ boson have
larger branching ratios than leptonic decay modes,
by using the hadronic channel, we can reconstruct more $Z$ bosons and hence select more signal events when the same luminosity is assumed. Therefore, compared with the results shown in Tab.~\ref{tab:2lep_signif}, the significances in Tab.~\ref{tab:2jet_signif} are better for the same benchmark points.

\section{Experimental sensitivity}
\label{sec:sensi}

In this section, we will discuss the experimental sensitivity to those effective operators given in Sec.~\ref{sec:eff_op}. It is known that $\mathcal{S}=3$ gives the $3\sigma$ experimental reach,
as we only consider statistical uncertainties.
After applying the selection cuts in the charged leptonic and hadronic channels,
$3\sigma$ reaches in the $m_\chi$-$\Lambda$ plane are obtained
at $e^+e^-$ colliders with $\sqrt{s}=250~\GeV$, 500~GeV, and 1~TeV,
as shown in Fig.~\ref{fig:3sigma_Lamb}.
For the operators $\mathcal{O}_\mathrm{F1}$ and $\mathcal{O}_\mathrm{F2}$,
we assume $\Lambda=\Lambda_1=\Lambda_2$,
which will turn off the DM coupling to $Z\gamma$.
The integrated luminosities are all assumed to be $1000~\ifb$.

\begin{figure*}[!htbp]
\centering
\subfigure[~Operator $\mathcal{O}_\mathrm{F1}$, $\Lambda=\Lambda_1=\Lambda_2$.]
{\includegraphics[width=.4\textwidth]{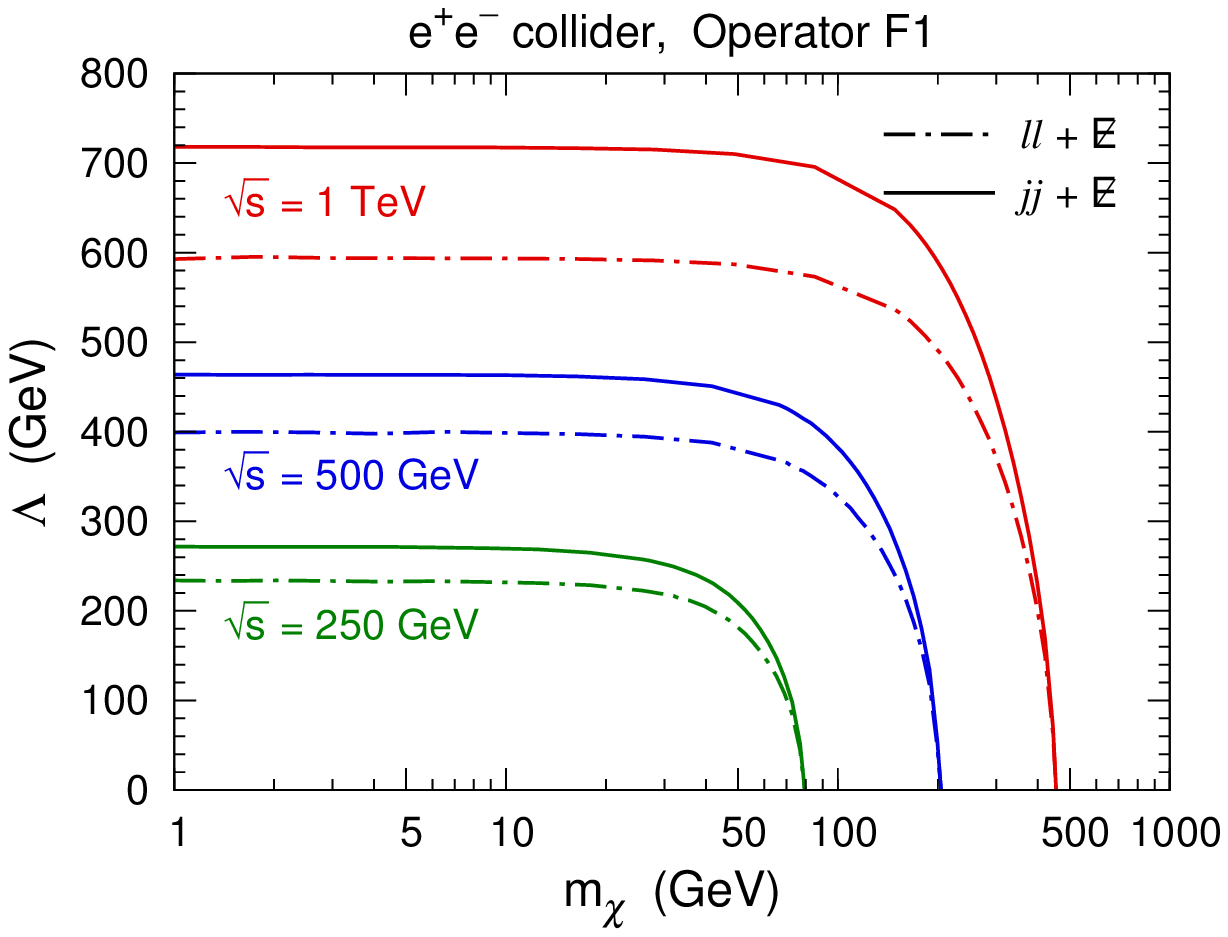}}
\subfigure[~Operator $\mathcal{O}_\mathrm{F2}$, $\Lambda=\Lambda_1=\Lambda_2$.]
{\includegraphics[width=.4\textwidth]{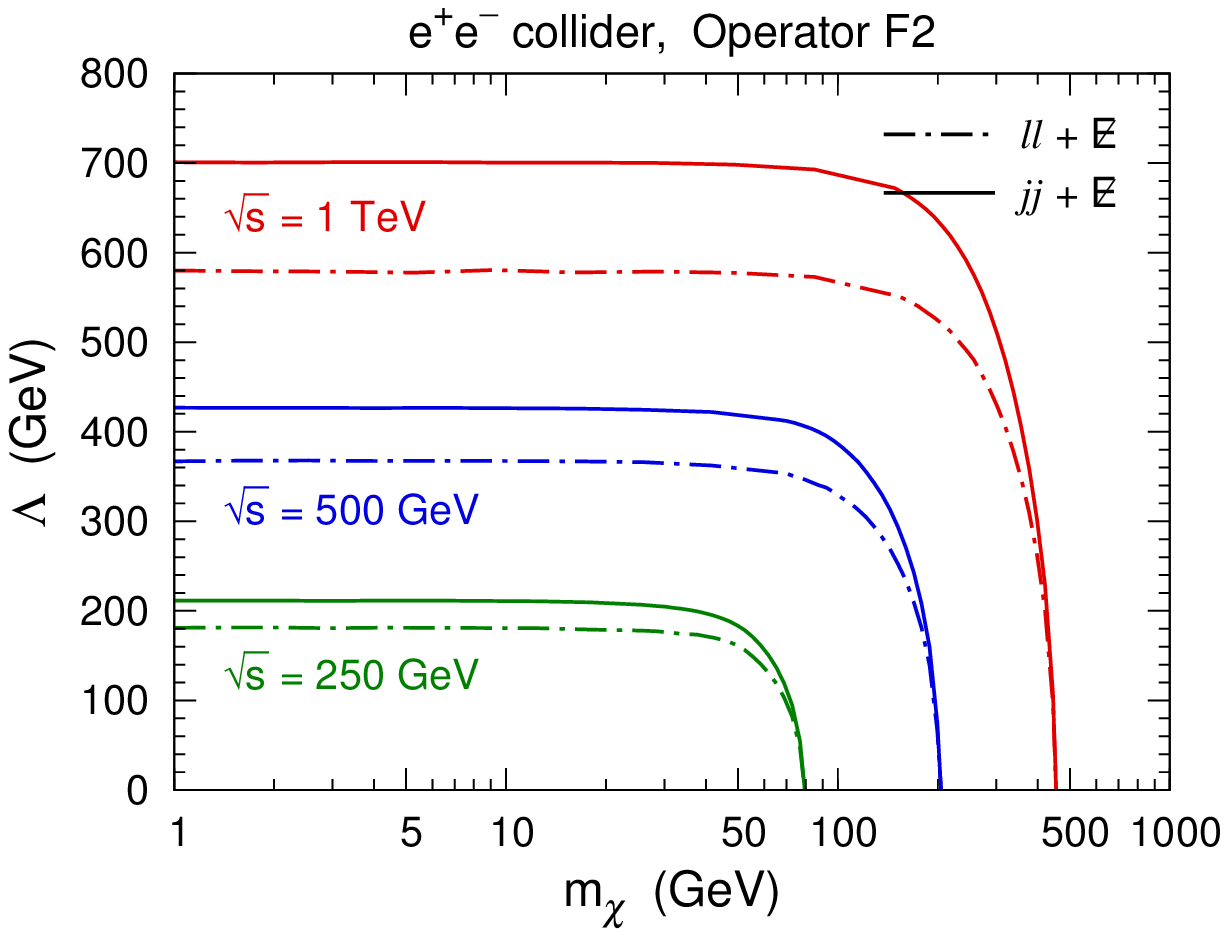}}
\subfigure[~Operator $\mathcal{O}_\mathrm{FH}$.]
{\includegraphics[width=.4\textwidth]{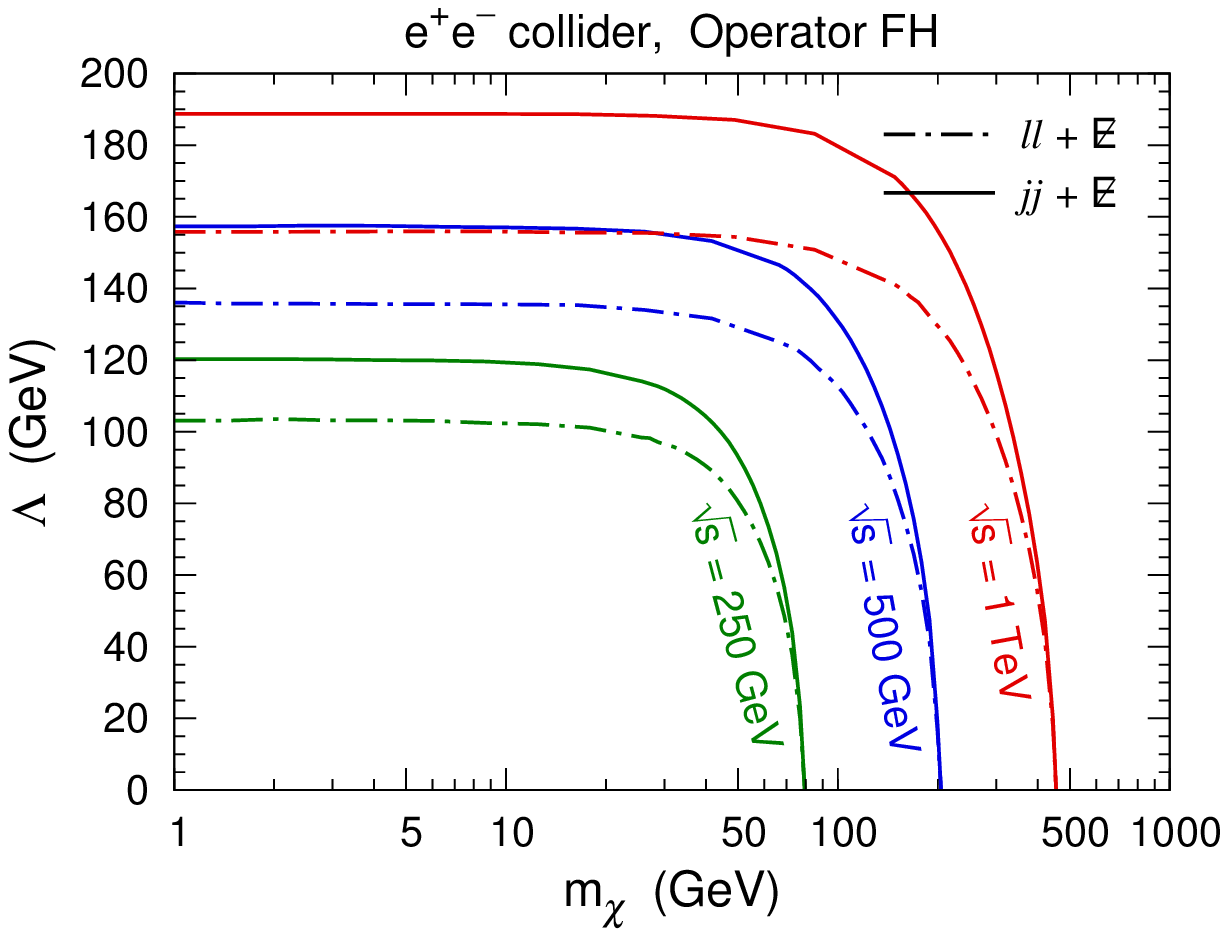}}
\subfigure[~Operator $\mathcal{O}_\mathrm{FP}$.]
{\includegraphics[width=.4\textwidth]{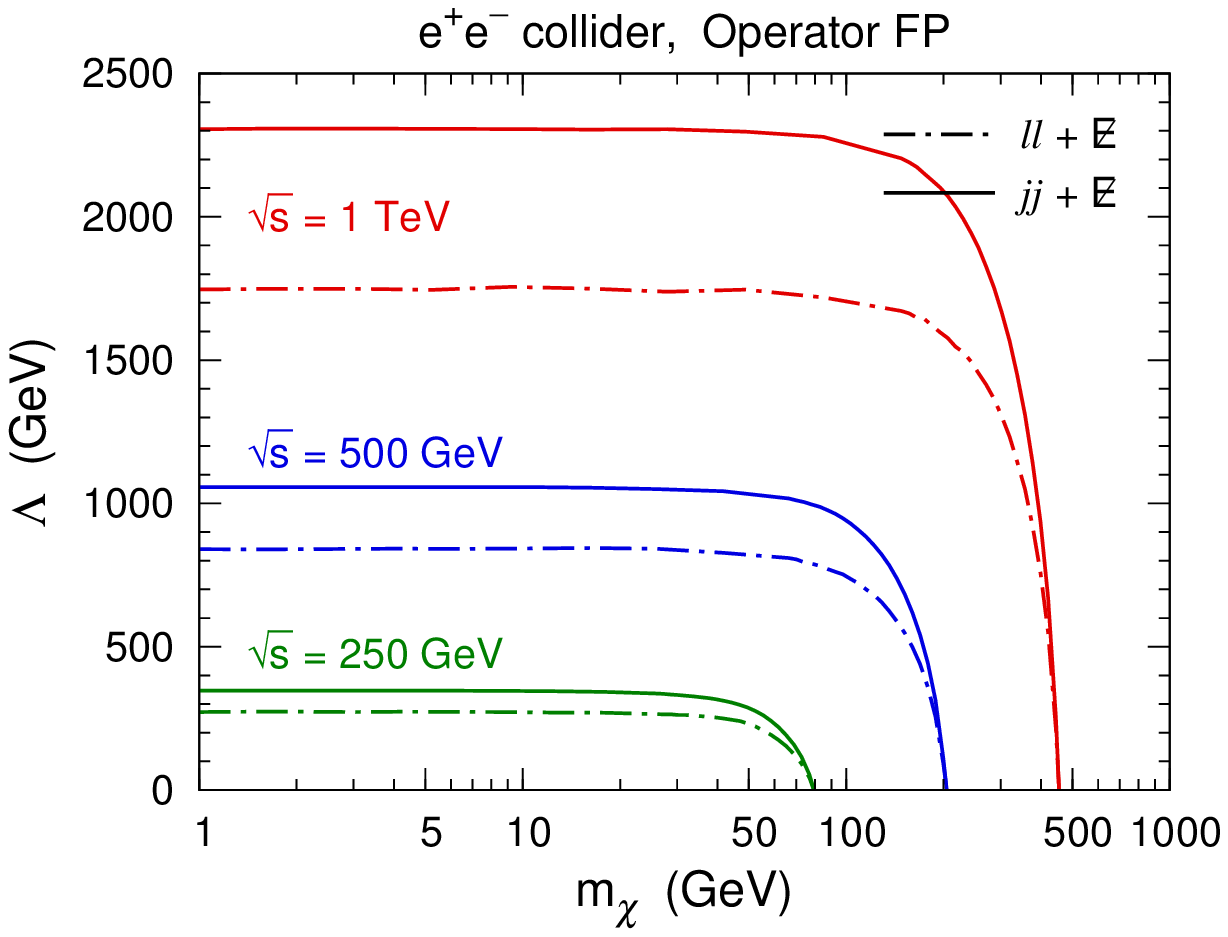}}
\subfigure[~Operator $\mathcal{O}_\mathrm{FA}$.]
{\includegraphics[width=.4\textwidth]{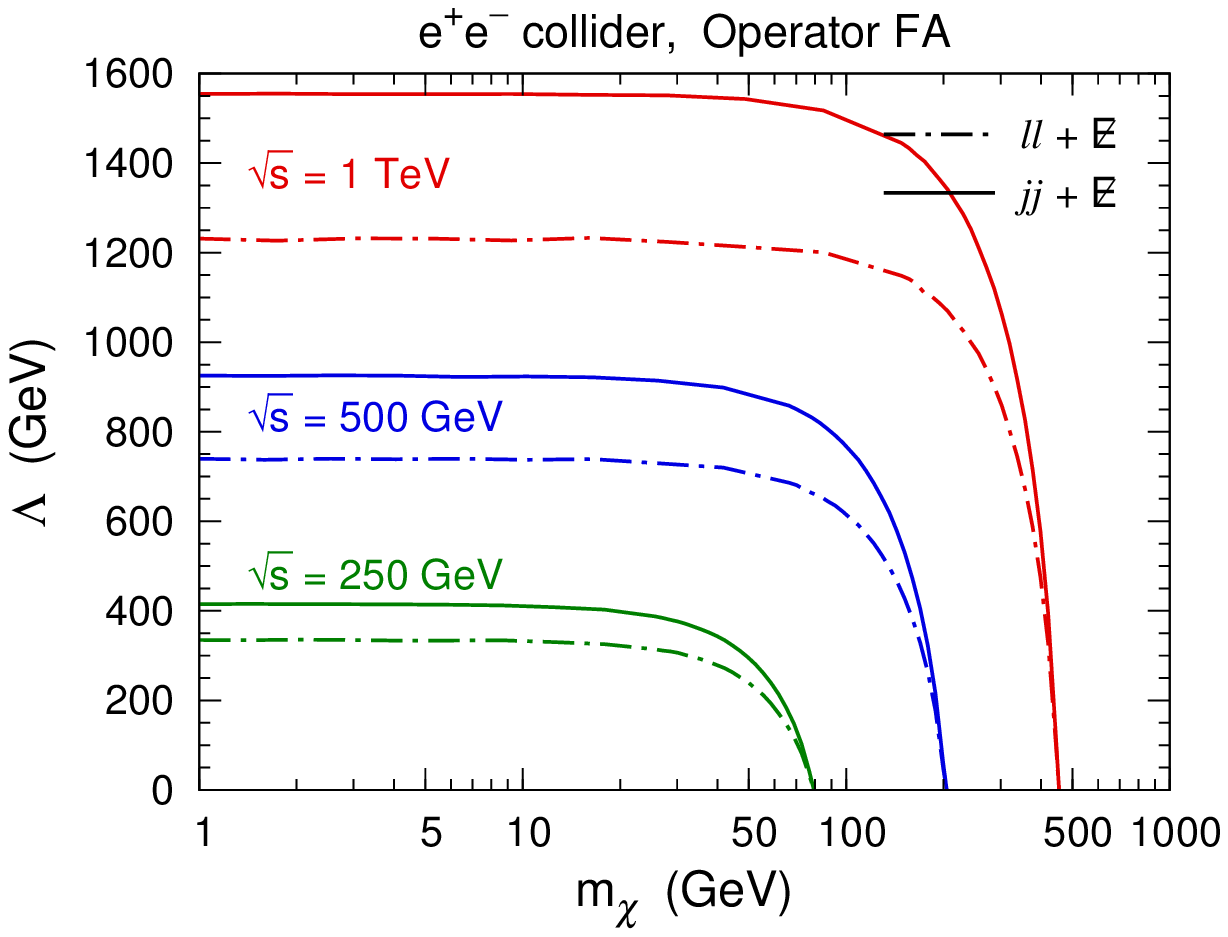}}
\caption{$3\sigma$ reaches in the $m_\chi$-$\Lambda$ plane
at $e^+e^-$ colliders with $\sqrt{s}=250~\GeV$, 500~GeV, and 1~TeV are provided.
The dot-dashed (solid) lines denote the charged leptonic (hadronic) channel.
Here the integrated luminosities are assumed to be $1000~\ifb$.}
\label{fig:3sigma_Lamb}
\end{figure*}

As we observed, the hadronic channel is more sensitive than the charged leptonic channel, as mentioned above. The experimental sensitivity decreases as the DM particle becomes heavy.
At the collision energy of 1~TeV, the operators $\mathcal{O}_\mathrm{F1}$ and
$\mathcal{O}_\mathrm{F2}$ can be explored up to $\Lambda\sim 700~\GeV$,
while the operator $\mathcal{O}_\mathrm{FH}$ can be explored only
up to $\Lambda\sim 190~\GeV$ due to the suppressed production cross section.
For the dimension-6 operators $\mathcal{O}_\mathrm{FP}$ and
$\mathcal{O}_\mathrm{FA}$, the $3\sigma$ sensitivities can reach up to
$\Lambda\sim 2.3~\TeV$ and $\Lambda\sim 1.6~\TeV$ at $\sqrt{s}=1~\TeV$, respectively.

It is interesting to compare our results with those estimated in Ref.~\cite{Chae:2012bq},
where the authors considered the DM production associating with
an initial state radiated \textit{photon}, $e^+e^-\to\chi\bar\chi\gamma$, for the operator $\mathcal{O}_\mathrm{FA}$.
We observe that our results are much less sensitive than theirs at $\sqrt{s}=250~\GeV$,
because the initial state radiated $Z$ boson is massive
and suppresses the phase space of the DM production cross section.
When the collision energy increases to 1~TeV,
however, our results are comparable with theirs,
since the $Z$ boson mass becomes negligible and the cross sections of these two processes are roughly equal at such a high energy.

\begin{figure*}[!htbp]
\centering
\subfigure[~Operator $\mathcal{O}_\mathrm{F2}$, $\Lambda_1=\Lambda_2$,
DM annihilation into $ZZ$.]
{\includegraphics[width=.4\textwidth]{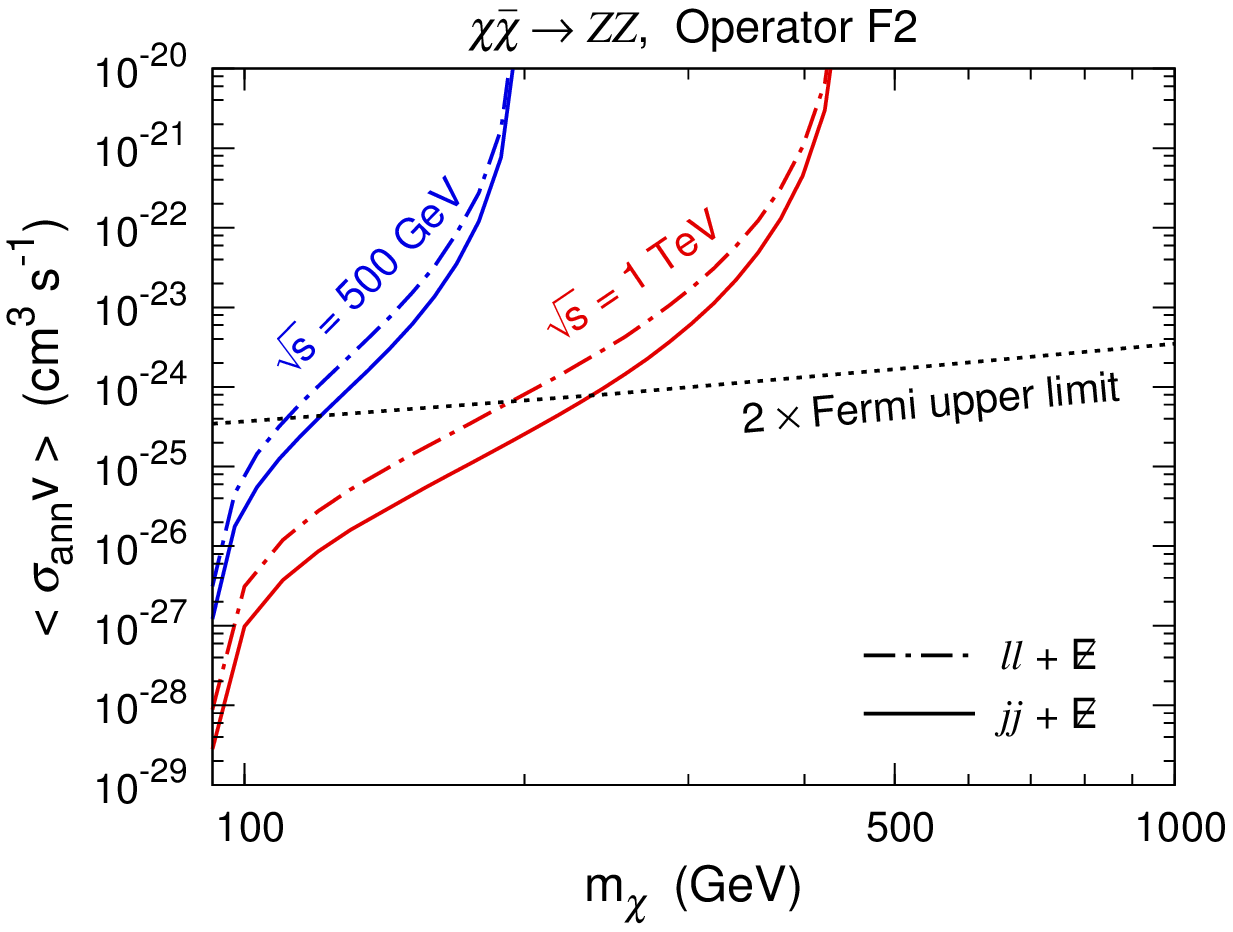}}
\subfigure[~Operator $\mathcal{O}_\mathrm{FP}$, DM annihilation into $e^+e^-$.]
{\includegraphics[width=.4\textwidth]{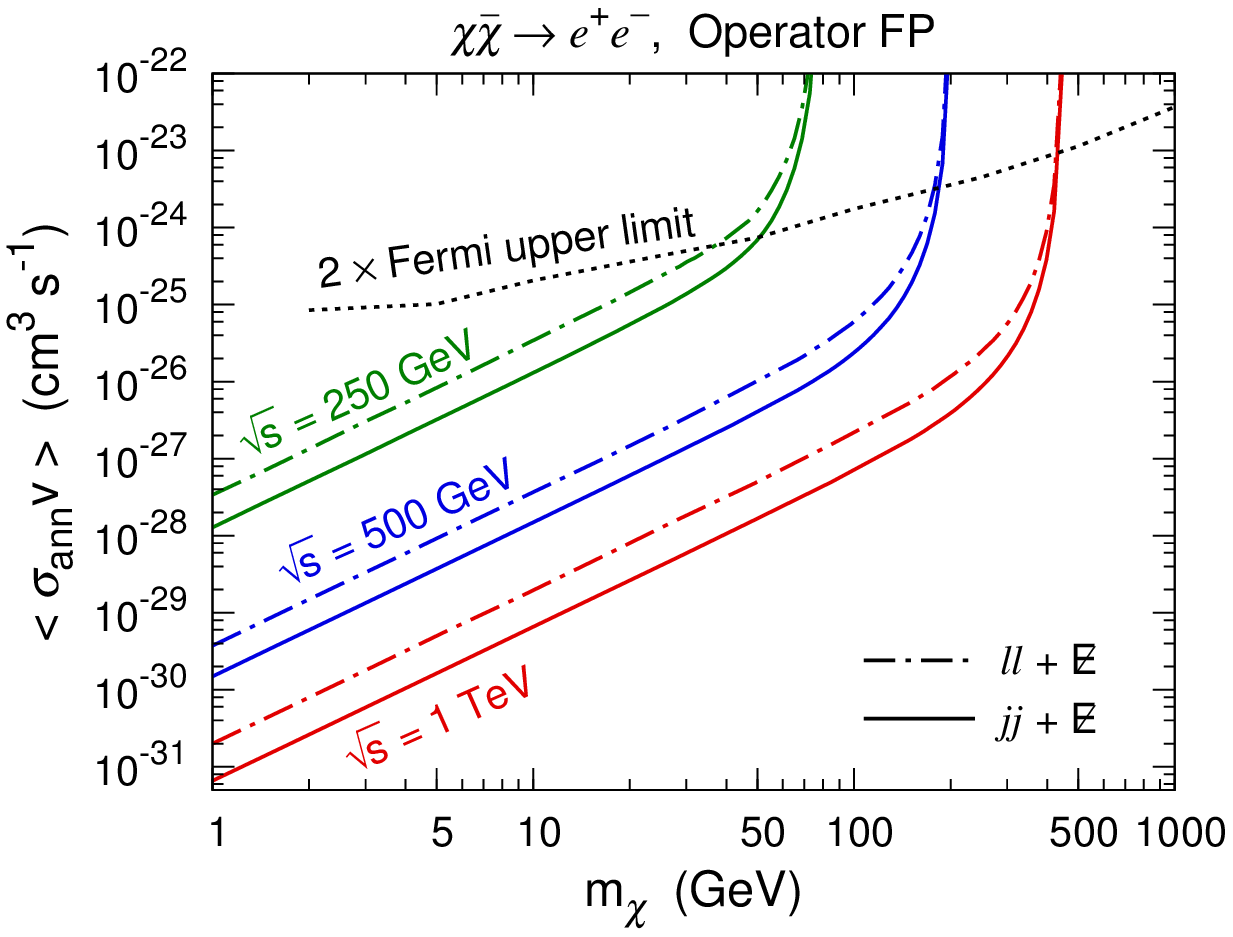}}
\caption{$3\sigma$ reaches in the $m_\chi$-$\left<\sigma_\mathrm{ann} v\right>$ plane
at $e^+e^-$ colliders with an integrated luminosity of $1000~\ifb$ are shown.
The dot-dashed (solid) lines denote the charged leptonic (hadronic) channel.
The dotted lines indicate the Fermi-LAT 95\% C.L. upper limits on
DM annihilation cross sections into $ZZ$ (left panel)
and into $e^+e^-$ (right panel)~\cite{Ackermann:2013yva}.
A factor of 2 is introduced to convert the Fermi-LAT limit from
self-conjugated DM particles to non-self-conjugated DM particles.}
\label{fig:3sigma_sv}
\end{figure*}

DM indirect searches can also be sensitive to DM couplings to $ZZ$ and to $e^+e^-$.
Based on 4-year $\gamma$-ray observations of 15 Milky Way
dwarf spheroidal satellite galaxies, the space experiment Fermi-LAT
have set 95\% C.L. upper limits on DM annihilation cross sections
into $W^+W^-$ and into $e^+e^-$~\cite{Ackermann:2013yva}.
Since the $\gamma$-ray spectrum yielded from the $ZZ$ channel is much similar to that yielded from the $W^+W^-$ channel,
we regard the limit on $\chi\bar\chi\to W^+W^-$ as the limit on $\chi\bar\chi\to ZZ$ once the kinematics is satisfied.

In the non-relativistic limit, the DM annihilation cross section into $ZZ$
due to the operator $\mathcal{O}_\mathrm{F2}$ can be expressed as~\cite{Chen:2013gya}
\begin{equation}
\left<\sigma_\mathrm{ann} v\right>_{\chi\bar\chi\to ZZ}
\simeq \frac{4m_\chi^4 G_\mathrm{ZZ}^2}{\pi}
\left(1 - \frac{m_Z^2}{m_\chi^2} \right)^{3/2},
\end{equation}
while the DM annihilation cross section into $e^+e^-$
due to the operator $\mathcal{O}_\mathrm{FP}$ is~\cite{Zheng:2010js}
\begin{equation}
\left<\sigma_\mathrm{ann} v\right>_{\chi\bar\chi\to e^+ e^-}
\simeq \frac{m_\chi^2}{2\pi \Lambda^4}\sqrt{1 - \frac{m_e^2}{m_\chi^2}}.
\end{equation}
Using these expressions, we convert the $3\sigma$ reaches
from the $m_\chi$-$\Lambda$ plane
into the $m_\chi$-$\left<\sigma_\mathrm{ann} v\right>$ plane,
as demonstrated in Fig.~\ref{fig:3sigma_sv}.
Note that the upper limits given in Ref.~\cite{Ackermann:2013yva} was assumed
that the DM particle and its antiparticle are identical.
In this work, however, we assume the DM particle to be a Dirac fermion,
and it is different to its antiparticle. The Fermi exclusion limit plotted
in Fig.~\ref{fig:3sigma_sv} has been scaled up by a factor of 2 in order to compensate this difference
(for a similar treatment, see Ref.~\cite{ATLAS:2012ky}).

Compared with indirect searches, collider searches are more sensitive to the mass of the DM particle. For the operator $\mathcal{O}_\mathrm{F2}$,
the $3\sigma$ reaches at an $e^+e^-$ collider with $\sqrt{s}=1~\TeV$
are more sensitive than the Fermi exclusion limit for $m_\chi\lesssim 200~\GeV$.
Moreover, for the operator $\mathcal{O}_\mathrm{FP}$, the $3\sigma$ sensitivity at $\sqrt{s}=1~\TeV$ can easily surpass the Fermi exclusion limit by several orders of magnitude for $m_\chi\lesssim 400~\GeV$.

On the other hand, DM annihilation processes due to the operators
$\mathcal{O}_\mathrm{F1}$, $\mathcal{O}_\mathrm{FH}$,
and $\mathcal{O}_\mathrm{FA}$ are either $p$-wave suppressed or helicity suppressed~\cite{Chen:2013gya,Kumar:2013iva}.
Indirect searches based on nonrelativistic dark matter in the universe are incapable of detecting them. Therefore, DM searches at colliders provide a unique way to probe the parameter space of these interactions.

\begin{figure}[!htbp]
\centering
\includegraphics[width=.4\textwidth]{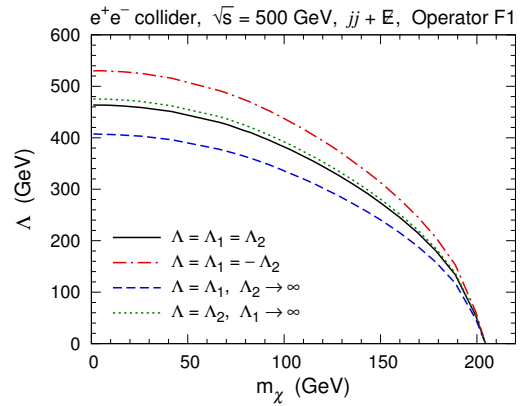}
\caption{$3\sigma$ reaches in the $m_\chi$-$\Lambda$ plane for different
$\Lambda_1$-$\Lambda_2$ relations in the hadronic channel at $\sqrt{s}=500~\GeV$ are shown with an integrated luminosity $1000~\ifb$.}
\label{fig:Lamb_compare}
\end{figure}

As indicated in Eqs.~\eqref{eq:G_ZZ} and \eqref{eq:G_AZ}, the DM couplings to
$ZZ$ and $Z\gamma$ for the operators $\mathcal{O}_\mathrm{F1}$
and $\mathcal{O}_\mathrm{F2}$ depend on two cutoff scales, $\Lambda_1$ and $\Lambda_2$.
The relation between $\Lambda_1$ and $\Lambda_2$ will determine
the relative contributions of the $\chi\chi ZZ$ and $\chi\chi Z\gamma$ couplings
to the DM production process $e^+e^-\to\chi\bar\chi Z$. As mentioned above,
$\Lambda_1=\Lambda_2$ will turn off the $\chi\chi Z\gamma$ coupling.
If $\Lambda_1=-\Lambda_2$ or $\Lambda_2\to\infty$, however,
the $\chi\chi Z\gamma$ coupling will be dominant.
If $\Lambda_1\to\infty$, the $\chi\chi ZZ$ and $\chi\chi Z\gamma$ couplings
will have comparable contributions.
These different $\Lambda_1$-$\Lambda_2$ relations
affect the DM production cross section and hence the signal significance,
as shown in Fig.~\ref{fig:Lamb_compare}.

\section{Polarized beams}
\label{sec:pol}

Polarized beams are sensitive to the chirality of underlying interactions.
At $e^+e^-$ colliders, the helicities of $e^-$ and $e^+$ are correlated by the spin of the particle exchanged in the $s$-channel.
Therefore, suitable polarization configurations may significantly enhance signal rates while efficiently suppress background processes.
Since new physics signal rates are typically predicted to be very small, making use of polarized beams can offer unique opportunities to discover them. In addition, the polarization of both beams makes it possible to perform unprecedented precise tests of the standard model, whose tiny deviations would be a hint for new physics.
For more benefits obtained from polarized beams, interested readers can refer Ref.~\cite{MoortgatPick:2005cw}.

In the baseline design of the ILC~\cite{Adolphsen:2013kya},
the electron (positron) source would be capable to provide a polarized beam
with a polarization degree of 80\% (30\%).
In this section, we will demonstrate how polarized beams benefit DM searches.
We take the case of $\sqrt{s}=500~\GeV$ as an illuminating example.

With longitudinal polarized beams, the cross section of a process
at an $e^+e^-$ collider can be expressed as~\cite{MoortgatPick:2005cw}
\begin{eqnarray}
\sigma(P_{e^-},P_{e^+})
&=& \frac{1}{4} \big[
 (1+P_{e^-})(1+P_{e^+}) \sigma_\mathrm{RR}
\nonumber\\
&&~ +(1-P_{e^-})(1-P_{e^+}) \sigma_\mathrm{LL}
\nonumber\\
&&~ +(1+P_{e^-})(1-P_{e^+}) \sigma_\mathrm{RL}
\nonumber\\
&&~ +(1-P_{e^-})(1+P_{e^+}) \sigma_\mathrm{LR}
\big],
\end{eqnarray}
where $P_{e^-}$ ($P_{e^+}$) is the polarization degree of the electron (positron) beam.
The right-handed (left-handed) polarization corresponds to
$P_{e^\pm}>0$ ($P_{e^\pm}<0$). The cross section for
the completely left-handed polarized $e^-$ beam ($P_{e^-}=-1$) and
completely right-handed polarized $e^+$ beam ($P_{e^+}=+1$) is denoted
as $\sigma_\mathrm{LR}$. Besides, $\sigma_\mathrm{LL}$, $\sigma_\mathrm{RR}$,
and $\sigma_\mathrm{RL}$ have analogous definitions.

\begin{table*}[!htbp]
\caption{Cross sections (in fb) for the main backgrounds and signals
under various polarization configurations after applying selection cuts are demonstrated.}
\label{tab:pol_Xsec}
\centering
\setlength\tabcolsep{0.4em}
\begin{minipage}{0.46\textwidth}
Charged leptonic channel, after cut~5 \\ \vspace*{0.2em}
\begin{tabular}{ccccccc}
\hline\hline
$(P_{e^-},P_{e^+})$ & $\ell^+\ell^-\bar\nu\nu$ & $\mathcal{O}_\mathrm{F1}$ & $\mathcal{O}_\mathrm{F2}$ & $\mathcal{O}_\mathrm{FH}$ & $\mathcal{O}_\mathrm{FP}$ & $\mathcal{O}_\mathrm{FA}$ \\
\hline
$(0,0)$ & 12.1 & 2.19 & 2.42 & 2.11 & 2.95 & 2.01 \\
$(+0.8,-0.3)$ & 2.02 & 2.19 & 2.43 & 2.08 & 2.24 & 1.97 \\
$(+0.8,+0.3)$ & 3.48 & 1.42 & 1.56 & 1.40 & 3.63 & 1.30 \\
$(-0.8,-0.3)$ & 15.2 & 1.89 & 2.09 & 1.82 & 3.63 & 1.75 \\
$(-0.8,+0.3)$ & 27.6 & 3.24 & 3.59 & 3.13 & 2.24 & 3.02 \\
\hline\hline
\end{tabular}
\end{minipage}
\begin{minipage}{0.46\textwidth}
Hadronic channel, after cut~5 \\ \vspace*{0.2em}
\begin{tabular}{cccccccc}
\hline\hline
$(P_{e^-},P_{e^+})$ & $jj\nu\bar\nu$ & $jj\ell\nu$ & $\mathcal{O}_\mathrm{F1}$ & $\mathcal{O}_\mathrm{F2}$ & $\mathcal{O}_\mathrm{FH}$ & $\mathcal{O}_\mathrm{FP}$ & $\mathcal{O}_\mathrm{FA}$ \\
\hline
$(0,0)$ & 92.6 & 2.91 & 15.1 & 17.1 & 14.1 & 20.1 & 12.9 \\
$(+0.8,-0.3)$ & 17.4 & 0.776 & 15.0 & 17.0 & 14.2 & 15.2 & 12.8 \\
$(+0.8,+0.3)$ & 26.2 & 1.07 & 9.77 & 11.2 & 9.17 & 24.9 & 8.40 \\
$(-0.8,-0.3)$ & 115 & 3.85 & 13.1 & 14.8 & 12.2 & 24.9 & 11.2 \\
$(-0.8,+0.3)$ & 212 & 6.05 & 22.3 & 25.3 & 20.9 & 15.2 & 19.0 \\
\hline\hline
\end{tabular}
\end{minipage}
\end{table*}

\begin{table*}[!htbp]
\caption{Signal significances with unpolarized beams ($\mathcal{S}_\mathrm{unpol}$)
and with optimal polarized beams ($\mathcal{S}_\mathrm{pol}$)
for the benchmark points at $\sqrt{s}=500~\GeV$ are compared with an integrated luminosity $100~\ifb$.}
\label{tab:pol_signif}
\centering
\setlength\tabcolsep{0.4em}
\begin{minipage}{0.28\textwidth}
\centering
Charged leptonic channel \\ \vspace*{0.2em}
\begin{tabular}{cccc}
\hline\hline
 & $\mathcal{S}_\mathrm{unpol}$ & $\mathcal{S}_\mathrm{pol}$ & $\mathcal{S}_\mathrm{pol}/\mathcal{S}_\mathrm{unpol}$ \\
\hline
$\mathcal{O}_\mathrm{F1}$ & 5.69 & 10.1 & 1.78 \\
$\mathcal{O}_\mathrm{F2}$ & 6.24 & 10.9 & 1.75 \\
$\mathcal{O}_\mathrm{FH}$ & 5.50 & 9.70 & 1.76 \\
$\mathcal{O}_\mathrm{FP}$ & 7.47 & 13.4 & 1.79 \\
$\mathcal{O}_\mathrm{FA}$ & 5.25 & 9.29 & 1.77 \\
\hline\hline
\end{tabular}
\end{minipage}
\begin{minipage}{0.28\textwidth}
\centering
Hadronic channel \\ \vspace*{0.2em}
\begin{tabular}{cccc}
\hline\hline
 & $\mathcal{S}_\mathrm{unpol}$ & $\mathcal{S}_\mathrm{pol}$ & $\mathcal{S}_\mathrm{pol}/\mathcal{S}_\mathrm{unpol}$ \\
\hline
$\mathcal{O}_\mathrm{F1}$ & 14.3 & 26.0 & 1.82 \\
$\mathcal{O}_\mathrm{F2}$ & 16.1 & 28.6 & 1.78 \\
$\mathcal{O}_\mathrm{FH}$ & 13.5 & 24.8 & 1.84 \\
$\mathcal{O}_\mathrm{FP}$ & 18.7 & 34.4 & 1.84 \\
$\mathcal{O}_\mathrm{FA}$ & 12.3 & 23.0 & 1.87 \\
\hline\hline
\end{tabular}
\end{minipage}
\end{table*}

In Tab.~\ref{tab:pol_Xsec}, we tabulate the cross sections
for the main backgrounds and signals under various polarization configurations
after applying selection cuts. The polarized cross sections of the main backgrounds
are primarily determined by the $e^\pm$ couplings to $W$ and $Z$ bosons.
In the standard model, $W^\pm$ can only couple to the left-handed $e^-$
and the right-handed $e^+$. $Z$ couples to $e^\pm$ via
$\dfrac{g_2}{2\cos\theta_W}(g_L \bar e_L \gamma^\mu e_L + g_R \bar e_R \gamma^\mu e_R)Z_\mu$,
where $g_L = -1 + 2\sin^2\theta_W \simeq -0.56$ and $g_R = 2\sin^2\theta \simeq 0.44$.
Thus the left-handed $e^-$ (right-handed $e^+$) coupling to $Z$ is stronger.
As a result, for the polarization configurations listed in Tab.~\ref{tab:pol_Xsec},
the backgrounds $\ell^+\ell^-\bar\nu\nu$,
$jj\nu\bar\nu$, and $jj\ell\nu$ are most suppressed with
$(P_{e^-},P_{e^+})=(+0.8,-0.3)$ and $(+0.8,+0.3)$.

For the operators $\mathcal{O}_\mathrm{F1}$, $\mathcal{O}_\mathrm{F2}$,
$\mathcal{O}_\mathrm{FH}$, and $\mathcal{O}_\mathrm{FA}$,
the completely polarized cross sections
$\sigma_\mathrm{LL}$ and $\sigma_\mathrm{RR}$ vanish,
due to the requirement of the vector and axial vector current interactions.
$\sigma_\mathrm{LR}$ is larger than $\sigma_\mathrm{RL}$
due to the stronger coupling between $Z$ and left-handed $e^-$.
However, the backgrounds with $(P_{e^-},P_{e^+})=(-0.8,+0.3)$ are very huge.
The optimal polarization configuration will be $(P_{e^-},P_{e^+})=(+0.8,-0.3)$.

On the other hand, for the operator $\mathcal{O}_\mathrm{FP}$,
$\sigma_\mathrm{LR}$ and $\sigma_\mathrm{RL}$ vanish,
because the pseudo-scalar interaction requires
the electron and positron have the same helicity.
$\sigma_\mathrm{LL}$ and $\sigma_\mathrm{RR}$ are equal.
In order to avoid huge backgrounds,
the optimal polarization configuration will be $(P_{e^-},P_{e^+})=(+0.8,+0.3)$.

\begin{figure*}[!htbp]
\centering
\subfigure[~Operator $\mathcal{O}_\mathrm{F1}$, $\Lambda=\Lambda_1=\Lambda_2$.]
{\includegraphics[width=.4\textwidth]{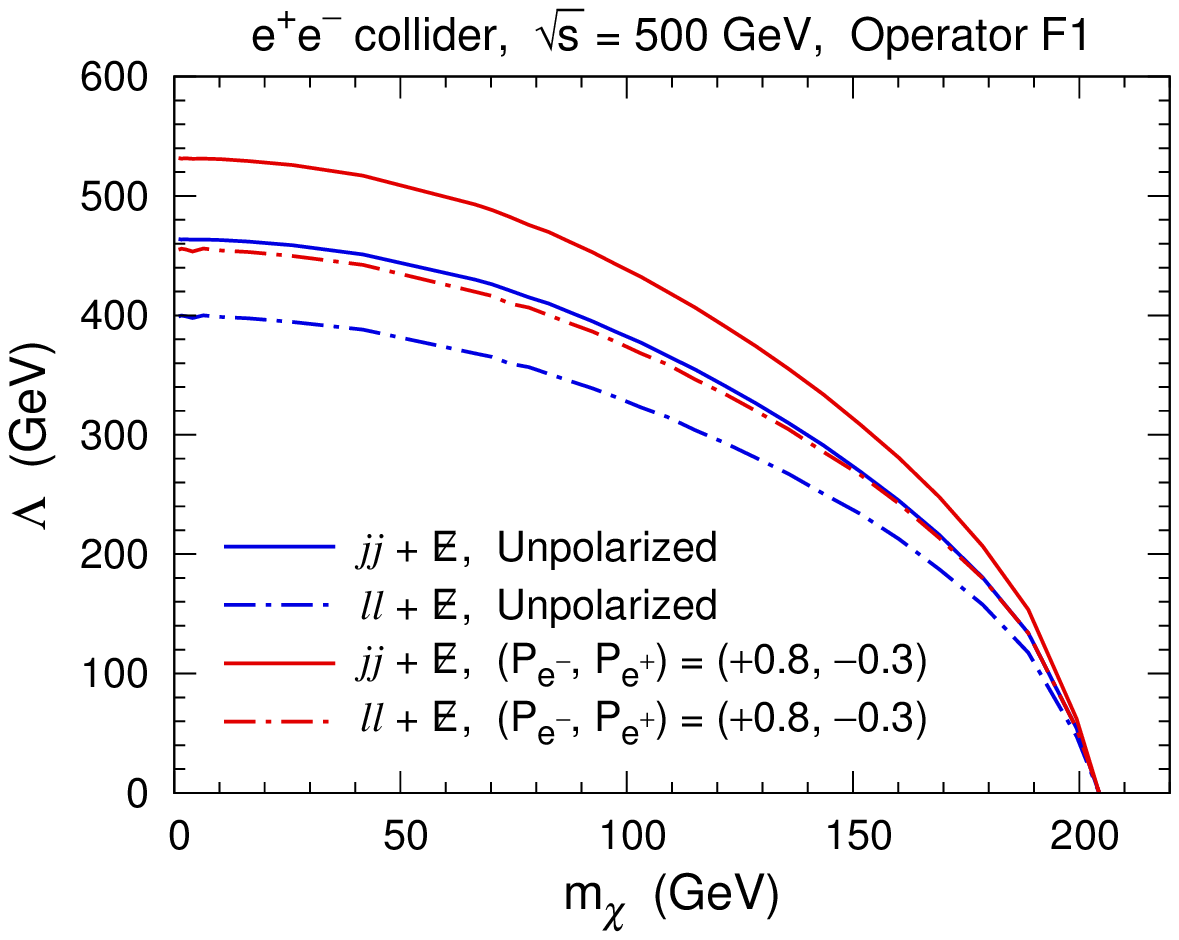}}
\subfigure[~Operator $\mathcal{O}_\mathrm{F2}$, $\Lambda=\Lambda_1=\Lambda_2$.]
{\includegraphics[width=.4\textwidth]{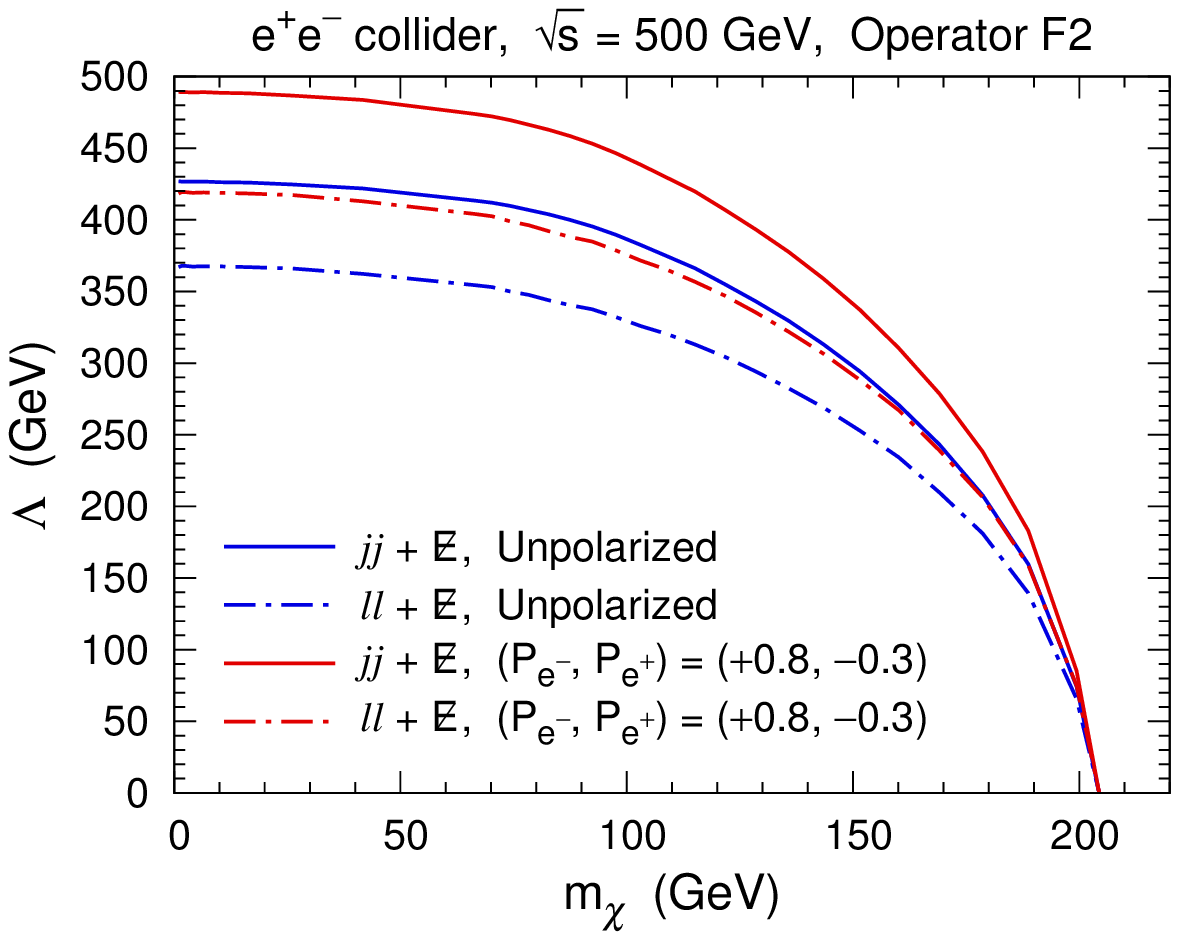}}
\subfigure[~Operator $\mathcal{O}_\mathrm{FH}$.]
{\includegraphics[width=.4\textwidth]{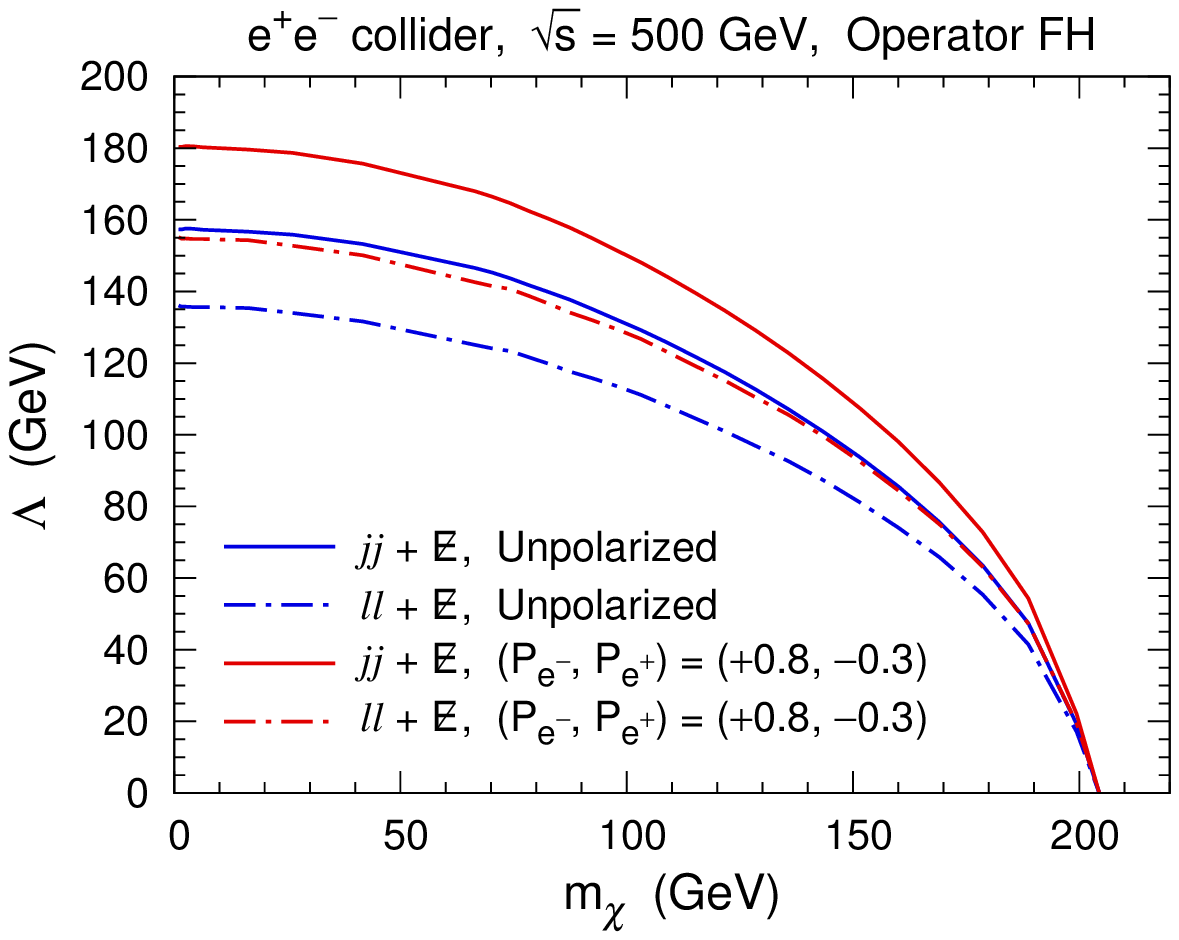}}
\subfigure[~Operator $\mathcal{O}_\mathrm{FP}$.]
{\includegraphics[width=.4\textwidth]{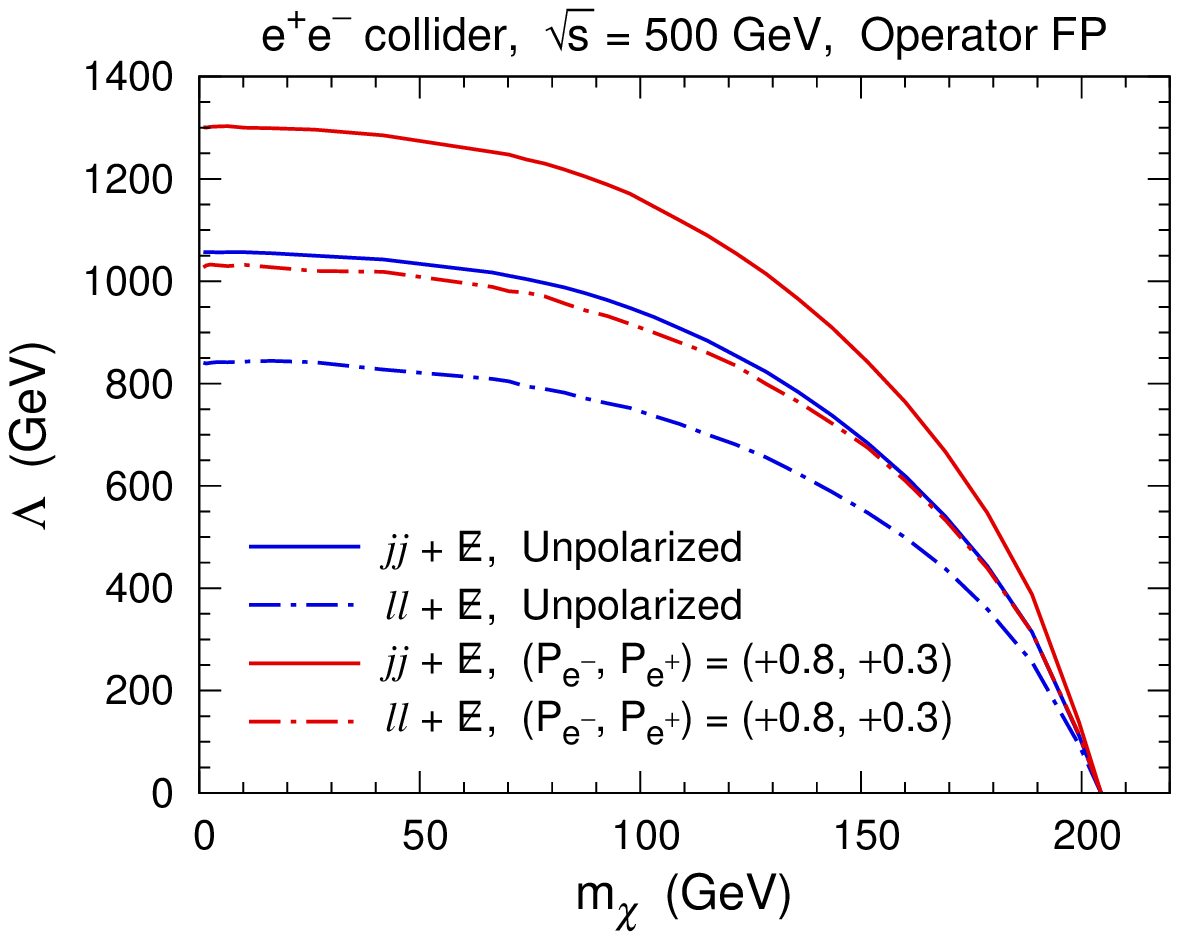}}
\subfigure[~Operator $\mathcal{O}_\mathrm{FA}$.]
{\includegraphics[width=.4\textwidth]{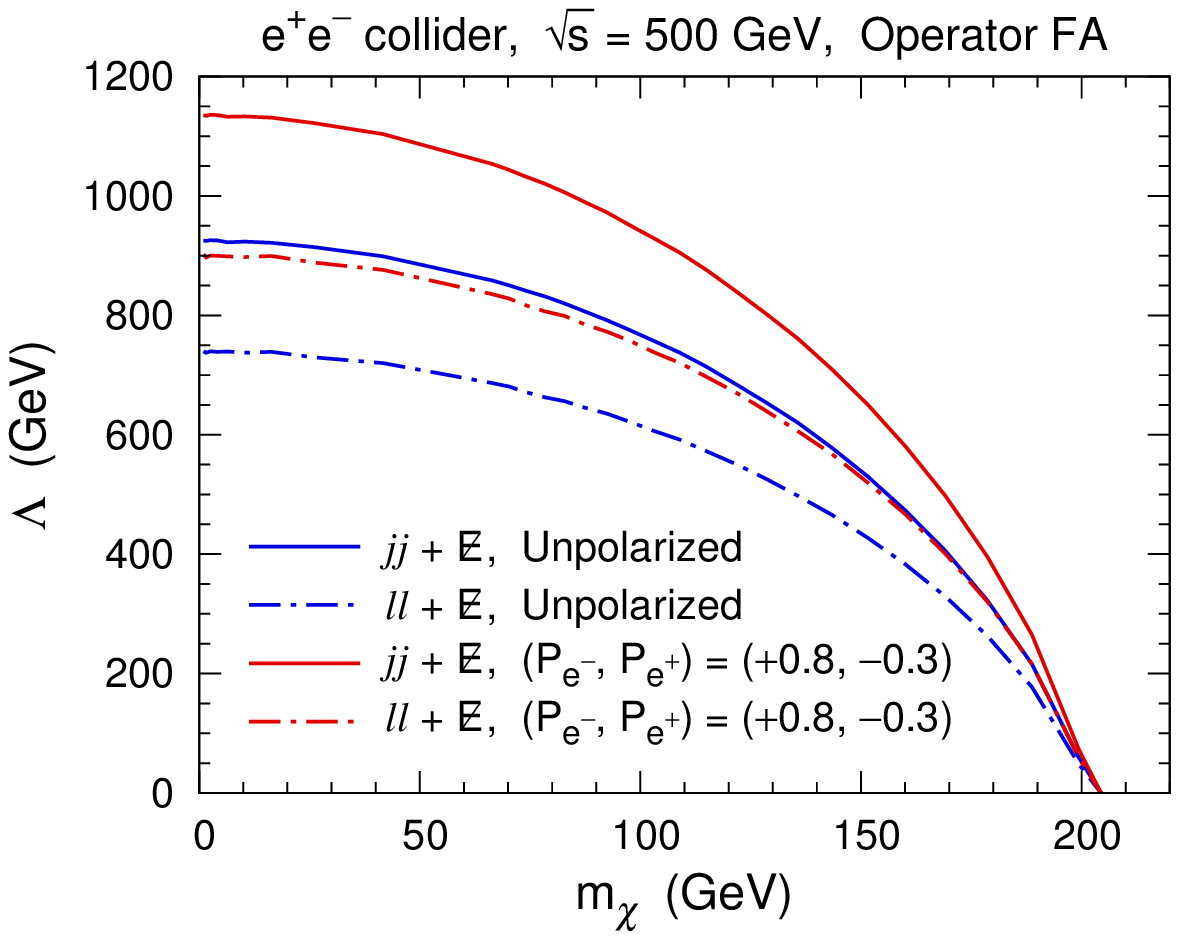}}
\caption{$3\sigma$ reaches in the $m_\chi$-$\Lambda$ plane
with unpolarized beams and with optimal polarized beams
in the charged leptonic (dot-dashed lines) and hadronic (solid lines) channels
for $\sqrt{s}=500~\GeV$ are shown with the integrated luminosity $1000~\ifb$.}
\label{fig:3sigma_Lamb_pol}
\end{figure*}

With an integrated luminosity of $100~\ifb$ and $\sqrt{s}=500~\GeV$,
signal significances with unpolarized beams ($\mathcal{S}_\mathrm{unpol}$)
and with optimal polarized beams ($\mathcal{S}_\mathrm{pol}$)
for the benchmark points are tabulated in Tab.~\ref{tab:pol_signif}.
A gain in significance is observed and the ratios $\mathcal{S}_\mathrm{pol}/\mathcal{S}_\mathrm{unpol}$
are in the range of $1.7-1.9$ in both the charged leptonic and hadronic channels.
Since the signal significance is basically proportional to the square root
of the integrated luminosity, using optimal polarized beams is
roughly equivalent to collecting more than three times of data.
In Fig.~\ref{fig:3sigma_Lamb_pol}, we show the $3\sigma$ reaches
in the $m_\chi$-$\Lambda$ plane with and without polarized beams
for $\sqrt{s}=500~\GeV$ and an integrated luminosity of $1000~\ifb$, where
the gains in significance are vividly demonstrated.

\section{Conclusions and discussions}
\label{sec:concl}

In this work, we focus on the dark matter searching sensitivity
in the mono-$Z$ channel at $e^+e^-$ colliders where $Z$ bosons can be reconstructed by either two charged leptons or two jets.
By using the cut-based method, we propose two sets of detailed kinematic cuts that can be used
in the charged leptonic and hadronic channels, respectively. It is observed that both the reconstructed $Z$ boson mass and the recoil mass against the $Z$ boson are crucial to distinguish
signal and background.
In the context of effective operators, we obtain the expected sensitivities
to the DM couplings to $ZZ/Z\gamma$ and to $e^\pm$.
Due to the larger branching ratios of the hadronic $Z$ decay modes and
relatively clean background at $e^+e^-$ colliders,
with the current analysis we observe that the significance in the hadronic channel is
better than that in the charged leptonic channel.

Compared with DM indirect searches by the Fermi-LAT experiment,
DM searches at $e^+e^-$ colliders have better sensitivity to the $\chi\chi ZZ$
and $\chi\chi ee$ couplings when the mass of the DM particle is below several hundred GeV.
Indirect searches for nonrelativistic DM may be incapable of
detecting some particular kinds of DM interactions due to $p$-wave or helicity suppressed annihilation cross sections.
In contrast, collider searches can cover such parameter space and can offer complementary searches to these interactions.

We emphasize that polarized beams can play an impor tant role for the DM searches.
We point out that in order to achieve the optimal sensitivity, different polarization configurations are needed to target different kinds of DM interactions, as demonstrated in Tab.~\ref{tab:pol_Xsec}. We also observed that using optimal polarized beams is roughly equivalent to acquiring more than three times of the integrated luminosity for the case of $\sqrt{s}=500~\GeV$.

\textit{Note added.} As this paper was in preparation,
a similar paper appeared~\cite{Neng:2014mga}, where
the searching for dark matter in the mono-$Z$ channel at $e^+e^-$ colliders is studied.
Nonetheless, it restricts to the study of interactions between DM particles and $e^\pm$.

\begin{acknowledgments}
Z.-H.~Yu thanks T.~Tait for helpful discussions. This work is supported by the Natural Science Foundation of China under Grants No.~11105157, No.~11175251, and No.~11135009, and the 973 project under Grant No. 2010CB833000.

\end{acknowledgments}


\end{document}